\documentclass[12pt]{iopart}
\usepackage{iopams}
\eqnobysec
\usepackage{epsfig}
\newcommand{\sump}{\mathop{{\sum}'}_{n=0}^\infty}
\renewcommand{\Re}{\mbox{Re }}

\begin{document}
\title[Nonsmoothness of the boundary]{Nonsmoothness of the boundary
and the relevant heat kernel coefficients}

\author{V~V~Nesterenko\dag\ \footnote[2]{To whom correspondence should
be addressed (nestr@thsun1.jinr.ru)}, I~G~Pirozhenko\dag\ and
J~Dittrich\S\ }

\address{\dag\ Bogoliubov Laboratory of Theoretical Physics,
Joint Institute for Nuclear Research,  141980 Dubna,  Russia}

\address{\S\ Physics Institute, ASCR, CZ-250 68 \v{R}e\v{z},
Czech Republic and Doppler Institute for Mathematical Physics,
Faculty of Nuclear Sciences and Physical Engineering, Czech
Technical University, Prague, Czech Republic}
\eads{\mailto{nestr@thsun1.jinr.ru},
\mailto{pirozhen@thsun1.jinr.ru}, \mailto{dittrich@ujf.cas.cz}}

\begin{abstract}
The contributions to the heat kernel coefficients generated by the
corners of the boundary are studied. For this purpose the internal
and external sectors of a wedge and a cone are considered. These
sectors are obtained by introducing, inside the wedge, a cylindrical
boundary. Transition to a cone is accomplished by identification
of the wedge sides.  The basic result of the  paper  is the
calculation of the individual contributions to the heat kernel
coefficients generated by the boundary  singularities. In the
course of this analysis certain patterns, that are followed by
these contributions, are revealed. The implications of the
obtained results in calculations of the vacuum energy for regions
with nonsmooth boundary are discussed. The rules for obtaining all
the heat kernel coefficients for the minus  Laplace operator
defined on a polygon or in its cylindrical generalization are
formulated.
\end{abstract}


\submitto{\CQG}

\maketitle

\section{Introduction}
     The asymptotic expansion of the heat kernel proves to be
important in a series of physical applications. Its coefficients
specify the divergences and conformal anomalies taking place in a
concrete field theory model~\cite{Od}, the high temperature
behaviour of the thermodynamic  functions~\cite{DowKen,BNP} and so
on~\cite{Kirsten}.

     For a flat manifold with a smooth boundary the heat kernel
coefficients are determined by local characteristics of the
boundary\ \cite{SW,Kennedy,ER,Gilkey}. If the boundary has
discontinuities (for example, it is piecewise smooth), then the
latter  give additional contributions to the heat kernel
coefficients\ \cite{AD1}. Usually in physical applications one
assumes that the boundary is smooth. However there is a series of
problems where such assumption is certainly not acceptable. A
typical example here is supplied by fields defined between two
plates which cross at a given angle, i.e., inside a dihedral
angle. Such a configuration is considered when calculating the
Casimir effect for a conducting wedge\ \cite{NLS}. If the fields
inside a dihedral are subjected to periodicity condition with
respect to the angular variable then one is concerned in fact with
the fields on  a cone, and the point where the boundary has
discontinuity becomes an internal point of the cone surface.
However the origin of this singularity is the same as in the case
of fields inside the wedge. The conical singularity proved to be
very important in many areas of mathematical
physics~\cite{Sommer,HowTr2} and lately it is investigated in
connection with  studies of quantum fields on the background of
black holes~\cite{ZCV} and cosmic strings~\cite{FS}.

     The general consideration of the boundary nonsmothness in
terms of the heat kernel expansion lacks till now\ \cite{Dowker}.
Such contributions to the heat kernel coefficients $C_{3/2}$ and
$C_2$ have been investigated in papers~\cite{AD1}. For a plane
domain  the contribution to $C_1$ generated by an edge of the
boundary is known\ \cite{Kac,MS} and by its limiting
configuration, by cusp. It is interesting to note that a cusp
pointing outward, with respect to the domain under study, leads to
change of  the power of the asymptotic variable (time) in the heat
kernel expansion \cite{SW} in comparison with the standard case.
In references\ \cite{SW,Kac,Pl,W} the asymptotic expansion of the
heat kernel with allowance for the boundary nonsmothness has been
built by calculating the relevant Green function of the heat
conduction equation. The present paper seeks to show the
effectiveness of applying the spectral zeta functions for the
calculation of the contributions  to  the heat kernel coefficients
caused by such boundary discontinuities as the corners. For this
goal we shall use the technique for constructing the zeta
functions developed in~\cite{Bordag} and extended in~\cite{BKD}.
This method proves to be very effective for calculating  the heat
kernel coefficients for different boundary conditions given on a
sphere and cylinder~\cite{BNP,Bordag,37}. A close approach
has been used in \cite{Moss,Esposito}.

     We shall consider internal and external parts of a plane
sector formed by two infinite radial rays emerging from the center
of a circle of radius $R$ at angle $\alpha $ to one another (see
figure 1, where I is the internal circular sector and II is the
external circular sector). The choice of such domains with
nonsmooth boundary is caused by the possibility of constructing
for them the global zeta functions. The latter cannot be done, for
example, for an open angle (figure~1 without circular arc inside
the angle). This point will be discussed in detail in Section~II.
In both the sectors the Laplace operator is defined   acting on
scalar real functions subjected to the Dirichlet or Neumann
boundary conditions.  If we substitute in figure~1 the radial rays
by crossed planes and the circular arc 1 -- 2 by an appropriate
part of a cylinder surface we arrive at the boundary value
problems which have the same heat kernel coefficients as in the
plane case.

  Identifying the points of the boundary with the polar
coordinates $(r, \theta=0)$ and $(r, \theta=\alpha)$, i.e.,
imposing the periodicity condition with respect to the angular
variable $\theta $ with a period $\alpha $,  we arrive obviously
at the spectral problem for the Laplace operator on two parts of
lateral surface of a cone $C_\alpha $: internal part $(r\le R)$
and external part $(r\ge R)$. At $r=R$ we can, as before, impose
the Dirichlet or Neumann conditions.

   In the present paper, six  coefficients of the asymptotic
expansion of the heat kernel  for the  boundary value problems
specified above will be calculated by making use of the relevant
zeta functions. It will be shown that each of these coefficients,
starting from the third one, is the sum of contributions generated
by the corners of the boundary and by the curvature of the arc
1--2. Analysis of the obtained results enables one to reveal some
regularities obeyed by the contributions of the boundary
nonsmoothness to the heat kernel coefficients, namely: i) the
corner contribution substantially depends on whether the corner is
made up by crossing two straight lines or two lines with nonzero
curvature; ii) contributions of adjacent angles to the heat kernel
coefficients $B_{3/2}$ and  $B_{5/2}$ have opposite signs in the
same way as the contributions of a circular arc to the
coefficients $B_1$ and $B_2$ for the internal and external
regions. In the case of a polygon or its cylindrical
generalization the rules are formulated for obtaining all the heat
kernel coefficients for the minus Laplace operator. The
implications of the obtained results in the  Casimir energy
calculations with employment of the zeta function technique are
also considered. In particular, it is shown that nonsmoothness of
the boundary does not  make always worse the situation with
calculation of the vacuum energy in the framework of the zeta
regularization method. For example, the corner contributions can
simply be mutually cancelled in the same way as the contributions
of the curved boundary are cancelled when taking into account the
internal and external regions. Besides, the analysis conducted
enables one to infer that it is very unlikely to get a finite and
unique result for the vacuum energy by smoothing the boundary
singularities, e.g., by taking into account the atomic structure
of the boundary or quantum fluctuations of the boundary.

     The paper is organized as follows.  Section 2 is devoted to
the detailed discussion of the choice of the domains with
piece-wise smooth boundary (internal and external circular
sectors), for which the complete spectral zeta functions can be
constructed. In section~3 the heat kernel coefficients are
calculated for internal circular sector by making use of the
counter integral representation for the corresponding zeta
functions \cite{Bordag,Kirsten}. In section~4 the contributions to
the heat kernel coefficients of the individual boundary
discontinuities (corners of different angles) are identified.
Further (in section 5) the heat kernel coefficients for the union
of both the sectors are calculated and the coefficients for the
external sector are obtained as the regarding differences.
Technically it turns out to be simpler in comparison with
calculation of the heat kernel coefficients  for the external
sector alone. The asymptotic expansion of the heat kernel for the
union of internal and external sectors is constructed by
differentiation of the logarithm of the function specifying the
frequency equation\ \cite{Moss,Esposito}. In section  6
the identification of the individual contributions of boundary
nonsmoothness to heat kernel coefficients for external sector is
carried out. Some general rules for these contributions are
revealed here for both sectors. In section~7 we conclude with a
few summarizing remarks.

\section{Choice of  domains with piecewise smooth boundaries}
When choosing the domain for investigating the contribution to the
heat kernel coefficients of the boundary discontinuities we pursue
two goals: the boundary of a domain should have a sufficient
number of discontinuities and at the same time for this domain one
can construct the spectral zeta function. In order to investigate
the corner singularities of the boundary  one should at first
sight take the most simple configuration, namely, the angle on
plane formed by two radial unrestricted rays or dihedral in space.
However, for these domains the global zeta functions cannot be
constructed. Let us explain this point in detail. We consider the
Laplace operator $\Delta $ acting on the scalar functions defined
inside the dihedral of opening angle $\alpha$ (wedge of an angle
$\alpha $, $W_\alpha\times {\mathbb R}^1$) and subjected to the
Dirichlet conditions on the wedge sides.  In cylindrical
coordinate system ($r,\theta,z$) the eigenfunctions in this
problem are
\[
u_{\lambda n k}(r,\theta, z)= \frac{\rme^{\rmi kz}}{\sqrt{\pi \alpha}} \,
 J_{np}(\lambda r)\sin (n p\, \theta){,}
\]
\begin{equation}
\label{eq2-1} 0\leq \theta\leq \alpha, \quad p=\pi /\alpha, \quad
n=1,2,\ldots , \quad 0\leq\lambda < \infty\,{,}
\end{equation}
where $J_\nu (z)$ is the Bessel function.
The operator $-\Delta $ has the following eigenvalues
\begin{equation}
\label{eq2-2}
\omega^2(k,\lambda)=k^2+\lambda^2,\quad -\infty <k<\infty,\quad 0\leq \lambda<
\infty\,{.}
\end{equation}
These eigenvalues do not depend on the quantum number $n$, i.e.,
there is a  degeneracy with respect to this number, the
multiplicity of this degeneracy being infinite\footnote{In
addition to infinite multiplicity, the every point of the spectrum
with fixed values of $k$ and $\lambda$ is a nonisolated point.
In fact, the spectrum is
continuous according to the terminology of spectral theory of operators
in Hilbert space~\cite{BEH,ReedS,BSh} and $\omega^2(k,\lambda)$ are not eigenvalues
as the eigenfunctions are not square integrable.}
\begin{equation}
\label{eq2-3}
N=\sum_{n=1}^{\infty}1=\infty \,{.}
\end{equation}

The global spectral zeta function of the operator $L$, $\zeta
(s)$, is determined as the trace of the ope\-ra\-tor~$L^{-s}$
\begin{equation}
\label{eq2-4}
\zeta (s) = \mbox{Tr }L^{-2s}=\sum_{j}N_j\lambda_j^{-2s}{,}
\end{equation}
where $\lambda_j^2$ is the $j$th eigenvalue of the operator $L$, and
$N_j$ is the degeneracy of this eigenvalue. Obviously,
this definition does not work
when $N_j$ is infinite.

In order to remove this degeneracy we put inside a dihedral a
cylindrical boundary as it is shown in figure~1 (the arc 1--2). On
the internal and external sides of this boundary the scalar field
will obey the Dirichlet or Neumann conditions. Thus, we are
considering the internal (I) and external (II)  sectors.
Certainly, the union of these two sectors is not identical to
unrestricted dihedral (to a wedge), because now the values of the
field on the arc 1--2 are not arbitrary but they are determined by
the corresponding boundary conditions. In the new configuration
there appear additional discontinuities of the boundary at the
points 1 and 2. However at these points the angle, at which the
involved boundary surfaces intersect, is fixed (it is equal to
$\pi/2$). It will be shown below that the contribution of such
boundary singularities to the heat kernel coefficients can be
easily separated from the contribution of the corner at the
origin. The latter is proportional to the difference $\pi
-\alpha$, because at $\alpha=\pi$ the singularity at the origin
disappears.

We shall concern with the standard
asymptotic expansion of heat kernel
\begin{equation}
\label{eq2-5}
K(t)=\sum_{j}e^{-\lambda_j^2t}=(4\pi
t)^{-d/2}\sum_{n=0}^{\infty}t^{n/2}
B_{n/2}+\mbox{ES},
\end{equation}
where $d$ is the dimension of the manifold under study and ES stands
for the exponentially small corrections as $t\to 0$.  This definition leads to
the same heat kernel coefficients $B_{n/2}$ for a dihedral and for
corresponding plane problem obtained by crossing of the dihedral by a
transverse plane. In fact, the eigenvalues of the operator $(-\Delta )$
in these two problems obey the relation
\begin{equation}
\label{eq2-6}
\lambda_{j}^2(d=3)=k^2+\lambda_{j}^2(d=2).
\end{equation}
Hence
\begin{equation}
\label{eq2-7}
K_{d=3}(t)=\int_{-\infty}^{\infty}\frac{\rmd k}{2\pi}\rme^{-k^2t}K_{d=2}(t)=
\frac{1}{\sqrt{4\pi t}}\,K_{d=2}(t){.}
\end{equation}
Taking into account the definition (\ref{eq2-5}) one easily deduce
from (\ref{eq2-7}) that  the heat kernel coefficients $B_{n/2}$ in
two eigenvalue boundary problems, mentioned above, are equal. Here
one should bear in mind that the heat kernel $K_{d=3}(t)$ and its
coefficients are referred to a unite length along the OZ axes. In
view of this, when calculating the  heat kernel coefficients we
shall consider either the spectral problem on a plane $(d=2)$ or
in the space $(d=3)$, pursuing only simplicity of calculation.

\section{Heat kernel coefficients for internal sector}
\subsection{Dirichlet boundary conditions}
At first we consider only internal sector (the region I in figure
1) for $d=3$, i.e., for internal sector of a dihedral or a
wedge\footnote{ In the framework of the quantum billiard studies
this problem has been investigated in~\cite{HowTr2}}. We employ
here the technique developed in~\cite{Bordag} and extended  to the
generalized bounded cone in~\cite{BKD} (see also
book~\cite{Kirsten}). In this approach the spectral zeta function
$\zeta(s)$ should be constructed in the beginning, and then the
relevant heat kernel coefficients are calculated through the
relation~\cite{BNP}
\begin{equation}
\label{eq3-1} \frac{B_n}{(4\pi)^{d/2}}=\lim_{s\to d/2-n}\left(
s+n-d/2 \right )      \zeta(s)\Gamma(s),\quad n=0,\,1/2,\,1,\,
\ldots\, {.}
\end{equation}

  For the Dirichlet boundary conditions the eigenfunctions of the
operator $(-\Delta)$ in the region I are defined by equation \
(\ref{eq2-1}), with $\lambda$ being the roots of the equations
\begin{equation}
\label{eq3-2}
J_{np}(\lambda_{nm}R)=0,\quad p=\pi/\alpha,\quad n=1,\,2,\,\ldots\,{.}
\end{equation}
   Here the subscript $m=1,\,2,\,\ldots $ numbers the nonzero roots
of these equations for fixed $n$. The relevant eigenvalues are
defined by equation\ (\ref{eq2-2}) with $\lambda=\lambda_{nm}$
\begin{equation}
\label{eq3-3}
\omega^2=k^2+\lambda^2_{mn}, \quad -\infty<k<\infty\,{.}
\end{equation}
In view of the behavior of the Bessel function near zero
$J_\nu(z)\sim z^\nu/(2^\nu\Gamma(\nu+1))$ it follows that the
frequency equation\ (\ref{eq3-2}) has the zero root of
`multiplicity' $\nu$.  Such roots should be removed from the
definition of the spectral zeta function (\ref{eq2-4}). Therefore
instead of equation\ (\ref{eq3-2}) we shall use the following
frequency equation
\begin{equation}
\label{eq3-4}
(\lambda R)^{-\nu} J_\nu(\lambda R)=0,\quad \nu=np, \quad n=1,\,2,\,
\ldots \,{.}
\end{equation}
  According to the general definition (\ref{eq2-4}) the zeta function
in the problem under study is given by
\begin{equation}
\label{eq3-5}
\zeta_D(s)=\int_{-\infty}^{\infty}\frac{\rmd k}{2\pi} \sum_{n=1}^{\infty}
\sum_{m}(k^2+\lambda_{nm}^2)^{-s}{.}
\end{equation}
As usual we substitute the sum over $m$ in this formula by the contour integral
in the plane of a complex variable~$\lambda$
\begin{equation}
\label{eq3-6}
\zeta_D(s)=\int_{-\infty}^{\infty}\frac{\rmd k}{2\pi}
\sum_{n=1}^{\infty}\frac{1}{2\pi i}\oint_C \rmd \lambda(\lambda^2+
k^2)^{-s}\frac{\rmd}{\rmd \lambda}
\ln [(\lambda R)^{-\nu} J_\nu(\lambda R)]\,{,}
\end{equation}
where the contour $C$  encloses counter clockwise the positive
roots of equations\ (\ref{eq3-4}). On deforming the contour $C$ in
an appropriate way and integrating with respect to $k$ by means of
the formula
\begin{equation}
\label{eq3-7}
\int_{-\infty}^{\infty}\frac{\rmd k}{(k^2+\lambda^2)^s}=\sqrt \pi
\lambda^{1-2s} \frac{\Gamma(s-1/2)}{\Gamma (s)},\quad \mbox{Re
}s>\frac{1}{2}{,}
\end{equation}
we arrive at the result
\begin{equation}
\label{eq3-8} \zeta_D(s)=\frac{\left ( R/p\right ) ^{2s-1}}{2
\sqrt \pi \Gamma (s) \Gamma\left (
\frac{3}{2}-s\right)}\sum_{n=1}^{\infty}n^{1-2s} \int_{0}^{\infty}
\rmd y\,y^{1-2s}\frac{\rmd}{\rmd y}\ln[(\nu y)^{-\nu} I_{\nu}(\nu y)]\,{,}
\end{equation}
where $I_\nu(z)$ is the modified Bessel function.

The analytical continuation of (\ref{eq3-8}) into the left
half-plane of the complex variable $s$ is accomplished by making
use of the uniform asymptotic expansion of the function $I_\nu(\nu
y)$
\begin{equation}
\label{eq3-9}
I_\nu(\nu y)\simeq \frac{1}{\sqrt{2\pi \nu}}\frac{\rme^{\nu
\eta}}{(1+y^2)^{1/4}}\left (
 1+\sum_{k=1}^{\infty}\frac{u_k(t)}{\nu^k}\right ){,}
 \end{equation}
where
 \begin{equation}
 \label{eq3-9a}
 \eta=\sqrt{1+y^2}+\ln\frac{y}{1+\sqrt{1+y^2}},\quad \frac{\rmd\eta
 }{\rmd y}=
 \frac{\sqrt{1+y^2}}{y}
 \,{,}
\end{equation}
and $u_k(t)$ are the known polynomials in $t=1/\sqrt{1+y^2}$.
Their explicit form and the corresponding recurrent relations can
be found in references\ \cite{AS,Olver}. Keeping in this expansion
all the terms proportional to $\nu^{-k}$ with $k\leq 4$  one can
write
\begin{equation}
\label{eq3-10}
\ln
\left (
 1+\sum_{k=1}^{\infty}\frac{u_k(t)}{\nu^k}\right )\simeq\sum_{k=1}^{4}
\frac{F^D_k(t)}{\nu^k}\,{,}
\end{equation}
where
\begin{eqnarray}
F^D_1(t)&=& \frac{1}{8}t -\frac{5}{24} t^3{,} \nonumber \\
F^D_2(t)&=&  \frac{1}{16}t^2-\frac{3}{8} t^4+\frac{5}{16}t^6{,} \nonumber \\
F^D_3(t)&=&{\frac {25}{384}}\,{t}^{3}-{\frac {531}{640}}\,{t}^{5}
+{\frac {221}{128}}\,{t}^{7}-{\frac {1105}{1152}}\,{t}^{9}{,}
              \nonumber \\
F^D_4(t)&=&
{\frac {13}{128}}\,{t}^{4}-{\frac {71}{32}}\,{t}^{6}+{\frac {531}{64}}
\,{t}^{8}-{\frac {339}{32}}\,{t}^{10}+{\frac {565}{128}}\,{t}^{12}{.}
\end{eqnarray}

In accordance with this expansion we represent the zeta function
in equation\ (\ref{eq3-9}) as the following sum
\begin{equation}
\label{eq3-11}
\zeta_D(s)=\sum_{j=-1}^{4}Z_j^D(s){,}
\end{equation}
where
\begin{eqnarray}
\fl
Z_{-1}^D(s)&=&C(s)p^{2-2s}\zeta_R(2s-2)\int_{0}^{\infty}\rmd y\,y^{1-2s}\frac{\rmd}{\rmd y}
(\eta-\ln y){,}\label{eq3-12} \\
\fl Z_{0}^D(s)&=&
-\frac{1}{4}\,C(s)p^{1-2s}\zeta_R(2s-1)\int_{0}^{\infty}\rmd y\,
y^{1-2s}\frac{\rmd}{\rmd y}\ln(1+y^2){,} \label{eq3-13} \\
\fl
Z_{j}^D(s)&=&C(s)p^{j-2s-1}\zeta_R(2s+j-1)\int_{0}^{\infty}\rmd y\,y^{1-2s}
\frac{\rmd}{\rmd y}F_j^D(t){,} \quad  j=1, 2,3,4{.} \label{eq3-14}
\end{eqnarray}
In these formulas $\zeta_R(s)$ is the Riemann  zeta function and
\begin{equation}
\label{eq3-15} C(s)= \frac{R^{2s-1}}{2\sqrt \pi \,\Gamma (s)\,
\Gamma (3/2-s)}{.}
\end{equation}

Substituting (\ref{eq3-9a}) into equation\ (\ref{eq3-12}) and
taking into account the value of the integral
\begin{equation}
\label{eq3-16} \int_{0}^{\infty}\rmd y
\frac{y^{2-2s}}{1+\sqrt{1+y^2}}=\frac{1}{4}\,
\frac{\Gamma(s-1)\,\Gamma(3/2-s)}{s-1/2}\,{,}
\end{equation}
we obtain for the function $Z_{-1}^D(s)$ the final expression
\begin{equation}
\label{eq3-17}
Z_{-1}^D(s)=\frac{R}{8\sqrt \pi}\left(\frac{R}{p}
\right
)^{2s-2}\frac{\Gamma(s-1)}{(s-1/2)\,\Gamma(s)}\,\zeta_R(2s-2)\,{.}
\end{equation}
The product $\Gamma (s)\,Z_{-1}^D(s)$ has simple poles at the points
$s=3/2,\; 1$, and $1/2$ with the respective residua
\begin{equation}
\label{eq3-18}
\frac{\alpha R^2}{16\pi^{3/2}}\,{,}\quad -\frac{R}{8\pi}\,{,}
\quad \frac{\sqrt{\pi}}{48\,\alpha}\,{.}
\end{equation}
Multiplying these residua by $(4\pi)^{3/2}$ we obtain, according
to equation\ (\ref{eq3-1}), the contributions of the function
$Z^D_{-1}(s)$ to the heat kernel coefficients $B_0,\,B_{1/2}$, and
$B_1$, respectively.

By making use of the table integral~\cite{GR}
\begin{equation}
\label{eq3-19}
2\int_{0}^{\infty} \frac{\rmd y\ y^{2-2s}}{1+y^2}=\Gamma \left (\frac{3}{2}-s
\right )\Gamma\left (s-\frac{1}{2}
\right ){,}
\end{equation}
we recast equation\ (\ref{eq3-13}) to the form
\begin{equation}
\label{eq3-20}
Z_0^D(s)=-\frac{1}{8\sqrt \pi} \left(\frac{R}{p}
\right )^{2s-1}\zeta_R(2s-1)\,\frac{\Gamma(s-1/2)}{\Gamma(s)}\,{.}
\end{equation}
The expression $\Gamma (s)\,Z_0^D(s)$ has simple poles at the points
$s=1/2$ and $s=1$ with the residua
\begin{equation}
\label{eq3-21}
-\frac{R}{16}\,\frac{\alpha}{\pi}\,{,} \quad \frac{1}{16\sqrt{\pi}}\,{,}
\end{equation}
which give the respective contributions to the heat kernel
coefficients $B_{1/2}$ and $B_1$ (see table 1).

Calculation of the functions $Z_j(s), \;j=1,\, \ldots \,, 4$,
defined in equation\ (\ref{eq3-14}), can be carried out with
hardly any trouble. Their contribution to the heat kernel
coefficients are given in table 1. In order to get the complete
values of these coefficients one should sum all the elements of
respective rows.
\subsection{Neumann boundary conditions}
In this case the operator $-\Delta$ has the following
eigenfunctions in internal circular sector I (see figure~1)
\[
v_{\lambda n k} (r, \theta,z)=\eta_{n0} \frac{\rme^{\rmi kz}}{\sqrt{\pi
\alpha}} \, J_{np}(\lambda_{nm} r) \,\cos (np \,\theta){,} \quad
n=0,1,2,\ldots ,\]
\begin{equation}
\label{eq3-22} 0\leq \theta\leq \alpha, \quad p=\pi \alpha, \quad
\eta _{n0}= \cases{\frac{{\displaystyle 1}}{{\displaystyle \sqrt
2}},\quad
 n=0,\cr 1,\quad  n=1,2,\ldots \, .\cr}
 \end{equation}
Here $\lambda_{nm}$ are the roots of the equations
\begin{equation}
\label{eq3-23}
J'_{np}(\lambda_{nm}R)=0, \quad n=0,\,1,\,2,\, \ldots \,{.}
\end{equation}
Taking into account the behavior of the derivative of the Bessel
function at the origin  $J'_\nu(z)\sim
z^{\nu-1}/(2^\nu\Gamma(\nu))$, we multiply equation\
(\ref{eq3-23}) by $(\lambda_{nm}R)^{1-np}$ in order to exclude the
zero multiple roots
\begin{equation}
\label{eq3-24}
(\lambda\, R)^{1-\nu}J'_\nu(\lambda R)=0,\quad \nu=np,\quad
n=0,\,1,\,\ldots\,{.}
\end{equation}

     By making  use of the frequency equations (\ref{eq3-24}) we can
write immediately the integral representations  for the zeta
function in the problem at hand analogous to equation\
(\ref{eq3-8})
\begin{eqnarray}
\label{eq3-25} \zeta_N(s)&=&C(s)\left
[\int_{0}^{\infty}\rmd y\,y^{1-2s}\frac{\rmd}{\rmd y}\ln (yI'_0(y))\right.
\\ && \nonumber \left.
+\sum_{n=1}^{\infty}n^{1-2s}\int_{0}^{\infty}\rmd y\,y^{1-2s}\frac{\rmd}{\rmd y}\ln
(y^{1-\nu}I'_\nu(\nu y)) \right ]{,}
\end{eqnarray}
where the coefficient $C(s)$ is has been defined in equation\
(\ref{eq3-15}). An important distinction of this equation is a new
term with $n=0$, that was absent in the equation\ (\ref{eq3-8})
for the Dirichlet boundary condition.

 We again use the uniform asymptotic expansion\ \cite{Olver,AS} but
now for the derivative of the Bessel function $I_\nu(\nu y)$
\begin{equation}
\label{eq3-26} I'_\nu(\nu y)\simeq \frac{1}{\sqrt{2 \pi
\nu}}\,\frac{(1+y^2)^{1/4}}{y}\,\rme^{\nu\eta}\left(
1+\sum_{k=1}^{\infty}\frac{v_k(t)}{\nu^k} \right ){,}
  \end{equation}
where $\eta$ is defined in equation\ (\ref{eq3-9a}) and the
functions $v_k(t)$ are the known polynomials \cite{AS} in $t(y)$.
Again we use the approximation
\begin{equation}
\label{eq3-27}
\ln
\left (
 1+\sum_{k=1}^{\infty}\frac{v_k(t)}{\nu^k}\right )\simeq\sum_{k=1}^{4}
\frac{F^N_k(t)}{\nu^k}\,{,}
\end{equation}
where
\begin{eqnarray}
F_1^N(t)&=&-\frac{3}{8}\,t+{\frac {7}{24}}\,{t}^{3},\nonumber \\
F_2^N(t)&=&-\frac{3}{16}\,{t}^{2}+\frac{5}{8}\,{t}^{4}-{\frac {7}{16}}\,{t}^{6}, \nonumber \\
F_3^N(t)&=&-{\frac {21}{128}}\,{t}^{3}+{\frac {869}{640}}\,{t}^{5}-{\frac {315}{
128}}\,{t}^{7}+{\frac {1463}{1152}}\,{t}^{9}, \nonumber \\
F_4^N(t)&=&-{\frac {27}{128}}\,{t}^{4}+{\frac {109}{32}}\,{t}^{6}-{\frac {733}{64
}}\,{t}^{8}+{\frac {441}{32}}\,{t}^{10}-{\frac {707}{128}}\,{t}^{12}{.}
\end{eqnarray}

    Instead of equation\ (\ref{eq3-11}) we have now
\begin{equation}
\label{eq3-29}
\zeta_{N}(s)= \tilde Z^N(s)+\sum_{j=-1}^{4}Z_j^N(s)\,{,}
\end{equation}
where
\begin{eqnarray}
\fl \tilde Z^N(s)&=&C(s)\int_{0}^{\infty} \rmd y
\,y^{1-2s}\frac{\rmd}{\rmd y}\ln
[yI_0'(y)]\,{,}\label{eq3-30} \\
\fl Z_{-1}^N(s)&=&Z_{-1}^D(s),\qquad   Z_{0}^N(s)=Z_{0}^D(s)\,{,} \label{eq3-30a}\\
\fl
Z_{j}^N(s)&=&C(s)\,p^{-2s-1+j}\zeta_R(2s+j-1)\int_{0}^{\infty}\rmd y\,y^{1-2s}
\frac{\rmd}{\rmd y}F_j^N(t){,}\quad  j=1, 2,3,4{.}  \label{eq3-30b}
\end{eqnarray}

Finding the functions $Z_j^N(s),\; j=1,\,\ldots,\,4$ in the
expansion (\ref{eq3-29}) presents no difficulty while the function
$\tilde Z^N(s)$ needs more thorough treatment.  By making use of the
asymptotics
\begin{eqnarray}
\frac{\rmd}{\rmd y}\ln[yI'_0(y)]&\simeq&\frac{2}{y}+\frac{1}{4}\,y+{\cal O}
(y^3) , \quad y\to 0\,{,} \label{eq3-31a} \\
\frac{\rmd}{\rmd y}\ln[yI'_0(y)]&\simeq&
1+\frac{1}{2y}+\frac{3}{8y^2}+\frac{3}{8 y^3}+\frac{63}{128y^4}+{\cal
O}(y^{-5}),\quad y\to \infty \label{eq3-31b}
\end{eqnarray}
it is easy to make sure that the integral in the definition of the
function $\tilde Z^N(s)$  does not converge at any values of $s$.
In order to overcome this drawback first the zeta function
$\zeta_N(s)$ for a scalar field with nonzero mass $m$ should be
constructed and on calculating the residua according to equation\
(\ref{eq3-1}) with this zeta function the mass $m$ should be put
equal to zero\footnote{Another way to treat $n=0$ case
for the Neumann boundary conditions has been explained in detail in~\cite{CEK}.}.
Following this line we consider the function
\begin{equation}
\label{eq3-32}
\tilde Z_m^N(s)=C(s)\int_{m}^{\infty} \rmd y \,( y^2-m^2
)^{-s+1/2}\frac{\rmd}{\rmd y}\ln [yI_0'(y)]\,{.}
\end{equation}
It is defined in the region $1<\Re s<3/2$, the lower (upper) limit in these
inequalities being determined by the convergence of the
integral when $y\to \infty$ ($y\to m$).

  In order to find, by equation\ (\ref{eq3-1}), the contribution of
the function $\tilde Z_m^N(s)$ to the coefficient $B_0$ the
analytical continuation of  $\tilde Z_m^N(s)$ to the
region $\Re s\geq 3/2$ is required. In the most simple way it can be
done by adding and subtracting from the integrand its
asymptotics when $y\to m$. For our goals it is sufficient to take only
the first  term in this asymptotics
\begin{equation}
\label{eq3-33} \fl
\frac{\tilde Z_m^N(s)}{C(s)}=\int_{m}^{\infty}\!\!\rmd y\,(y^2-m^2)^{1/2-s}\left
\{ \frac{\rmd}{\rmd y}\ln [yI'_0(y)] - f_m \right \}+
f_m\int_{m}^{\infty}\!\!\rmd y\, (y^2-m^2)^{1/2-s}{,}
\end{equation}
where
\[
f_m=\left .\frac{\rmd}{\rmd y}\ln[yI'_0(y)]\right |_{y=m}{.}
\]
The first term in equation\ (\ref{eq3-33}) is regular  at the
point $s=3/2$, but the integral in the second term gives a simple
pole at this point
\[
\int_{m}^{\infty}\!\! \rmd y \,(y^2-m^2)^{1/2-s}=\frac{m^{1-2s}}{2}\frac{\Gamma(s-1)\,
\Gamma(3/2-s)}{\Gamma(1/2)}{.}
\]
However the gamma function $\Gamma (3/2-s)$ responsible for this
pole is canceled by the same multiplier in the denominator of the
coefficient $C(s)$ (see equation\ (\ref{eq3-15})). As a result the
function $\tilde Z_m^N(s)$ does not give the contribution to the
heat kernel coefficient $B_0$.

     For calculating the contribution of the function $\tilde
Z_m^N(s)$ to the rest heat kernel coefficients  by  equation\
(\ref{eq3-1}), this function should be analytically continued to
the region $\Re s\leq 1$. It can again be done by adding and
subtracting the asymptotics of the integrand  when $y\to \infty $
now. Further the residua are found according to equation\
(\ref{eq3-1}) and only after that the mass $m$ is put equal to
zero. The corresponding results are presented in  table 2. Finding
the contributions of the rest functions
$Z_j^N(s),=\;j=-1,\,\ldots,\, 4$ to the heat kernel coefficients
presents no trouble (see table 2).

\subsection{Spectral problems on a cone}
Imposing on the eigenfunctions the periodicity condition with
respect to the angular variable $\theta$ with a period $\alpha$ we
pass from a wedge to a cone because in this case the respective
points on the radial rays $O1$ and $O2$ are identified (see
figure~1), the circular arc 1--2 being converted into a
circumference. Apparently, the boundary discontinuities at the
points 1 and 2 disappear. This circumference separates two parts
(internal and external) of the cone surface which for simplicity
will be refereed to as the internal and external sectors of the
cone.  On the circumference separating them we impose on the
eigenfunctions, as before, the Dirichlet or Neumann conditions.
The corresponding (unnormalized) eigenfunctions for internal cone
sector are
\begin{equation}
\label{eq3-34} u_{nm}(r,\theta)= J_{np}(\lambda_{nm}r)\left ({\sin
np\,\theta \atop\cos np\,\theta} \right ),\quad
p=\frac{2\pi}{\alpha},\quad n=0,\,1,\,2,\,\ldots\,{,}
 \end{equation}
where $\lambda_{nm}$ are the roots of the frequency equations
for the Dirichlet boundary conditions at $r=R$
\begin{equation}
\label{eq3-35}
J_{np}(\lambda_{nm}R)=0,\quad n=0,\,1,\,2,\, \ldots
\end{equation}
 or of those for the Neumann conditions
\begin{equation}
\label{eq3-36}
J'_{np}(\lambda_{nm}R)=0,\quad n=0,\,1,\,2,\, \ldots\,{.}
\end{equation}

For the external sector the Bessel functions in equations\
(\ref{eq3-34}) -- (\ref{eq3-36}) are replaced by the Hankel
functions $H^{(1)}_{np}(\lambda r)$.

From (\ref{eq3-34}) it follows that on the cone all the states
with $n\not= 0$ are double degenerate
\begin{equation}
\label{eq3-37}
N_0=1,\quad N_n=2, \quad n=1,\,2,\, \ldots\, {.}
\end{equation}

Thus, in order to find the heat kernel coefficients  for the internal
cone sector one should put in previous calculations
\begin{equation}
\label{eq3-38} p=2 \pi/\alpha\,{,}
\end{equation}
 take into account the degeneracy of states (\ref{eq3-38}), and sum
 up over $n$ starting with $n=0$.

     Let us consider the internal sector of a cone with the Dirichlet
conditions on the circumference 1--2. For the corresponding zeta
function $\zeta^c_D(s)$ the representation (\ref{eq3-8}) holds
with $p$ defined in (\ref{eq3-38}) and with the summation replaced
by $2\,\sump$, where the prime on the summation sign means that
the $n=0$ term is counted with half weight. For $\zeta_D^c(s)$ in
the sum (\ref{eq3-11}) there arises an additional term with $n=0$
\begin{equation}
\label{eq3-39}
\zeta_D^c(s) = \tilde  Z^D(s) + 2\sum_{j=-1}^{4}Z_j^D(s)\,{,}
\end{equation}
where
\begin{equation}
\label{eq3-40}
 \tilde  Z^D(s)=C(s)\int_{0}^{\infty}\rmd y \,y^{1-2s}\frac{\rmd}{\rmd y}
\ln I_0(y)\,{,}
\end{equation}
and the functions $Z_j^D(s),\;j=-1,\,0,\, \ldots\, ,4$ are defined
by equations\ (\ref{eq3-12}) -- (\ref{eq3-14}) with $p=2
\pi/\alpha$.

The asymptotics
\begin{eqnarray}
\frac{\rmd}{\rmd y}\ln I_0(y)&\simeq&\frac{y}{2}-\frac{y^3}{16}+{\cal O}
(y^5) , \quad y\to 0\,{,} \label{eq3-41} \\
\frac{\rmd}{\rmd y}\ln I_0(y)&\simeq&
1-\frac{1}{2y}-\frac{1}{8y^2}-\frac{1}{8 y^3}-\frac{25}{128y^4}+{\cal
O}(y^{-5}),\quad y\to \infty \label{eq3-42}
\end{eqnarray}
imply that the function  $ \tilde  Z^D(s)$ in equation\
(\ref{eq3-40}) is defined in the region
\begin{equation} 1< \Re s < 3/2.
\end{equation}
To single out in the integral (\ref{eq3-40}) the pole contribution
at the point $s=3/2$ we rewrite the function $\tilde Z^D(s) $ as
follows
\begin{eqnarray}
\tilde  Z^D(s) &=&C(s)\int_{0}^{1} \rmd y\,y^{1-2s}
\left [
\frac{\rmd}{\rmd y}\ln I_0(y) -\frac{y}{2}
\right ]
\nonumber \\
&&+\frac{C(s)}{2}\int_{0}^{1}
\rmd y\,y^{2-2s}+C(s)\int_{1}^{\infty}\rmd y\, y^{1-2s}\frac{\rmd}{\rmd y}I_0(y)\,{.}
\label{eq3-44}
\end{eqnarray}
The integrals in the first and third terms in equation\
(\ref{eq3-44}) are regular at the point $s=3/2$. Substituting the
second term from equation\ (\ref{eq3-44}) into definition
(\ref{eq3-1}) we get
\[
\mathop{\mbox{Res}}\limits_{s\to 3/2{-0}}(\tilde Z^D(s)\Gamma(s))=
\mathop{\mbox{Res}}\limits_{s\to 3/2{-0} }\left (
\frac{R^{2s-1}}{2\sqrt \pi\Gamma(3/2-s)}\frac{1}{3-2s} \right
)=0\,{.}
\]
     Thus the function $\tilde  Z^D(s) $ does not give any
contribution to the coefficient $B_0$.

     So as to find the contribution of the function $\tilde Z^D(s)
$ to the heat kernel coefficients $B_n,\; n=1/2,\,1,\, \ldots $ we
again split the domain of integration in equation\ (\ref{eq3-40})
into two intervals $(0,\,1)$ and $(1,\,\infty)$. When integrating
over the second interval we add and subtract under the integral
sign the asymptotics (\ref{eq3-42}). When calculating the residua
at the points $s=1,\;s=1/2,\;0,\;-1/2,\;-1$ we shall take the
right-hand limits in equation\ (\ref{eq3-1}). It gives the
following contributions to the heat kernel coefficients
$B_{1/2},\,B_1,\,B_{3/2},\,B_2$, and $B_{5/2}$, respectively
\begin{equation}
\label{eq3-45}
2\,R\sqrt\pi,\quad-\pi,\quad-\frac{\sqrt \pi}{2\,R},\quad
-\frac{\pi}{4R^2},\quad
-\frac{25\sqrt \pi}{48\,R^3}\,{.}
\end{equation}

In order to evaluate the contributions of the  functions
$Z_j^D(s),\; j=-1,\,0, \,\ldots,\, 4$ to the heat kernel
coefficients, in addition to (\ref{eq3-45}), one should substitute
in table 1 $\alpha $ by $\alpha /2$ and multiply all the elements
of this table  by 2. Summing the contributions (\ref{eq3-45}) and
the data from table 1 we obtain the heat kernel coefficients for
the internal sector I on the cone with the Dirichlet conditions on
the circle 1--2
\[
B_0=\frac{\alpha\,R^2}{2},\quad B_{1/2}=-\frac{1}{2}\,\alpha\,R\,
\sqrt \pi,\quad B_1=\frac{2}{3}\,\frac{\pi ^2}{\alpha}+\frac{\alpha}{6},
\]
\begin{equation}
\label{eq3-45a}
 B_{3/2}=\frac{\alpha\,\sqrt \pi}{64\,R},\quad B_{2}=\frac{4\,\alpha}{315\,R^2},
\quad  B_{5/2}=\frac{37\,\alpha\,\sqrt \pi}{2^{13}}\,{.}
\end{equation}

  For  Neumann boundary conditions the spectral zeta
function for internal sector on a cone is given by
\begin{equation}
\label{eq3-46} \zeta_N^c(s)=\tilde
Z^N(s)+2\sum_{j=-1}^{4}Z_j^N(s)\,{,}
\end{equation}
where the functions $\tilde Z^N$ and $Z_j^N$ are determined in
equations\ (\ref{eq3-30}) -- (\ref{eq3-30b}) with  $\alpha $
replaced by $\alpha/2$. The corresponding coefficients of the heat
kernel expansion are
\[
B_0=\frac{\alpha\,R^2}{2},\quad B_{1/2}=\frac{1}{2}\,\alpha\,R\,
\sqrt \pi,\quad B_1=\frac{2}{3}\,\frac{\pi ^2}{\alpha}+\frac{\alpha}{6},
\]
\begin{equation}
\label{eq3-46a}
 B_{3/2}=\frac{5\,\alpha\,\sqrt \pi}{64\,R},\quad B_{2}=\frac{4\,\alpha}{45\,R^2},
\quad  B_{5/2}=\frac{269\,\alpha\,\sqrt \pi}{2^{13}}\,{.}
\end{equation}
\section{Identification of the individual contributions to the heat
kernel coefficients for internal sector}
 Let us envisage the  internal sector I (see figure\ 1). Its boundary
possesses the following peculiarities which give contribution to
the heat kernel coefficients: nonzero curvature of the arc 1--2;
right-angled corners at the points 1 and 2; corner of angle
$\alpha$ at the origin. The arc contribution is proportional to
its length, i.e., to $\alpha$,  the contributions of the
right-angled corners does not depend on $\alpha $, contribution of
the corner at the origin vanishes when $\alpha =\pi$.  It is
sufficient to separate in each of the heat kernel coefficients the
contribution due to  each boundary  singularity enumerated above.
We demonstrate this considering the heat kernel coefficients for
internal circular sector I with Dirichlet conditions on its
boundary (see table 1).

The first coefficient
\begin{equation}
\label{eq4-a}
B_0=\alpha\,\frac{R^2}{2}=|\Omega|
\end{equation}
is the area $|\Omega|$ of the circular
sector I. The second heat kernel coefficient
\begin{equation}
\label{eq4-b}
B_{1/2}=-\sqrt \pi\,\frac{2\,R+\alpha\,R}{2}=-\sqrt
\pi\,\frac{L}{2}\,{,}
\end{equation} where $L$ is the length of the sector boundary (its
perimeter). As concerns the coefficient
\begin{equation}
\label{eq4-1} B_1=\frac{1}{6}\left (\frac{\pi^2}{\alpha}+\alpha
\right )+\frac{\pi}{2}\,{,}
\end{equation}
the situation is more complicated. It is clear that $\pi/2$ is the
contribution of two right-angled corners at the points 1 and 2.
The term $\alpha /6$  contains the contribution of the curvature
of the arc 1--2 (denote it by $k_{\rm{arc}}\,\alpha$) and a part
of the contribution of the corner at the origin $O$ (the latter is
equal to $(1/6-k_{\rm{arc}}) \,\alpha$). In terms of these
notations the complete contribution of the corner at the origin is
\begin{equation}
\label{eq4-2} c(\alpha)=\frac{\pi^2}{6\,\alpha}+\left
(\frac{1}{6}-k_{\rm{arc}} \right )\alpha\,{.}
\end{equation}
From the condition
\begin{equation}
\label{eq4-3} c(\pi)=\pi\left (\frac{1}{3}-k_{\rm{arc}} \right
)=0
\end{equation}
it follows that
\begin{equation}
\label{eq4-4} k_{\rm{arc}}=1/3\,{.}
\end{equation}
Further we find  $c(\alpha)$
\begin{equation}
\label{eq4-5}
c(\alpha)=\frac{\pi^2-\alpha^2}{6\,\alpha}\,{.}
\end{equation}
Thus the corner of an angle $\alpha$ on the boundary gives the
contribution to $B_1$ defined by equation\ (\ref{eq4-5}). At first
time this contribution has been calculated  in\
\cite{Kac}. Another method to derive it is described in\
\cite{MS}. In both the cases the Green's function of the equation
of heat conductivity was considered. In our approach it is found
by making use of the spectral zeta function technique.

     Now we are in position to check the consistency of our reasoning,
namely, we can calculate the contribution to the coefficient $B_1$
due to the right-angled corners at the points 1 and 2 by making
use of equation\ (\ref{eq4-5}). It gives
\begin{equation}
\label{eq4-6} 2\, c(\pi /2)=\pi/2\,{.}
\end{equation}
It is this value that has been attributed to this contribution
above (see equation\ (\ref{eq4-1})).

     Identification of individual contributions to the rest of heat
kernel coefficients can be done in a direct way. The terms
independent of the angle  $\alpha $ are attributed to the
right-angled corners at the points 1 and 2, while the linear in
$\alpha $ terms are due to the curvature of the arc 1--2 (see
 table 3). Merely such a separation of individual peculiarity
contributions leads to a correct value of the arc curvature
contribution  which is known from the heat kernel expansion for a
smooth boundary (see below). It is worth noting that the angle
$\alpha $ at the origin does not contribute to the heat kernel
coefficients $B_n$ with $n>1$ even when $\alpha=\pi/2$. It may be
explained only taking into account that the higher derivatives of
the radius vector of the boundary curve behave in a different way
at the origin and at the points 1 and 2.

     Let $\Omega $ be  a simply connected  region of a plane bounded
by a smooth  curve $\Gamma $. For the heat kernel of the minus
Laplace operator with the Dirichlet conditions on $\Gamma $ the
following asymptotic expansion holds when $t \to 0$ (see, for
example, references\ \cite{SW,Kac,MS})
\begin{eqnarray}
\fl
K(t)&\simeq&\frac{|\Omega|}{4\,\pi\,t}-\frac{L}{8\,\sqrt{\pi\,t}}+
\frac{1}{12\pi}
\int_{\Gamma}k(s)\,\rmd s+\frac{\sqrt{\pi\,t}}{256\,\pi}\int_{\Gamma}{k^2(s)\,\rmd s}
+\frac{t}{315\,\pi}\int_{\Gamma}k^3(s)\,\rmd s \nonumber \\
\fl
&&+\sqrt{\pi\,t^3}\left [
\frac{37}{2^{15}\pi}\int_{\Gamma}k^4(s)\,\rmd s-\frac{11}{2^{11}\pi}\int_{\Gamma}
(k'(s))^2\,\rmd s \right ]+{\cal O}(t^2)\,{,} \label{eq4-7}
\end{eqnarray}
where $|\Omega |$ is the area of $\Omega $, $L$ is the length of
$\Gamma $, $k(s)$ is  the curvature of the curve $\Gamma $ at the
point $s$, $k^2(s)=(\rmd^2{\bf r}/\rmd s^2)^2$, where ${\bf r}(s)$ is a
parametric representation of the curve $\Gamma $; $s$ is the natural
parameter along $\Gamma $: $\rmd s^2=(\rmd{\bf r})^2$; $k'(s)=\rmd k(s)/\rmd s$.
For convex portions of $\Gamma $ $k(s)$ is considered to be positive,
and for concave parts of $\Gamma $  $k(s)$ is assumed to be negative.

   In the expansion (\ref{eq4-7}) the numerical coefficients of
$k,\;k^2, \;k^3$, and $k^4$ are derived from the contributions,
proportional to $\alpha $, to the heat kernel coefficients
$B_1,\;B_{3/2},\;B_2$, and $B_{ 5/2}$, respectively (see  table 3).
Here it should be taken into account that in the problem under
study $k(s)=1/R,\;\rmd s=R\,\rmd\alpha$ and the coefficients $B_n$ enter
the heat kernel expansion (\ref{eq2-5}) with the multiplier
$1/(4\,\pi)$.

   If we go from a wedge to a cone by identifying the radial rays
$O1$ and $O2$  the corners at the points 1 and 2 disappear. The
sole singular point remains the origin $O$ which becomes an
internal point of the cone surface. However the contribution of
this singularity to the coefficient $B_1$ has the same nature as
in the case of wedge.

    The heat kernel coefficients  for internal sector I on the
cone with Dirichlet condition on the boundary 1--2 are listen in
equation\ (\ref{eq3-45a}). The coefficients $B_0$ and $B_{1/2}$
obey general equations (\ref{eq4-a}) and (\ref{eq4-b}) with
corresponding values for $|\Omega|$ and $L$. Separation of the
contributions to $B_1$ generated by singularity at the origin and
by the curvature of the circle 1--2 can be conducted in the same
way as for a wedge. Let $k_{{\rm  arc}}\,\alpha$  be the
contribution of the boundary 1--2 and
\begin{equation}
\label{eq4-11}
d(\alpha)=\frac{2}{3}\,\frac{\pi^2}{\alpha}+\frac{\alpha}{6}-
k_{{\rm arc}}\,\alpha
\end{equation}
     be the contribution due to the singularity at the origin. When
$\alpha=2\pi $ the surface of a cone becomes a plane and the
singularity at the origin $O$ disappear. Therefore, $d(2\,\pi)=0$. It
gives
\begin{equation}
\label{eq4-12} k_{{\rm arc}}=1/3\,{,}
\end{equation}
i.e., for this quantity we have  the same value as in the case of
a wedge (see  (\ref{eq4-4})). With (\ref{eq4-12}) allowed for, one
deduces from the equation\ (\ref{eq4-11})
\begin{equation}
\label{eq4-13}
 d(\alpha)=\frac{2}{3}\,\frac{\pi^2}{\alpha}-\frac{\alpha}{6}
=2\,c(\alpha/2)\,{,}
\end{equation}
where $c(\alpha)$ is defined in equation\ (\ref{eq4-5}).

     Identification of the individual contributions to the heat
kernel coefficients for the Neumann boundary conditions is
conducted in the same way (see  table 3). The first coefficient
$B_0$  is, as before, the area $|\Omega|$ of the sector I. For the
second coefficient $B_{1/2}$ wehave equation (\ref{eq4-b}) with
oposite sign in the right-hand side
\begin{equation}
\label{eq4-13a}
B_{1/2}=\sqrt \pi \,\frac{L}{2}, \quad L=2\,R+\alpha \,R\,{.}
\end{equation}
The contribution to the coefficient $B_1$ generated by the
singularity at the origin proves to be the same as for the
Dirichlet conditions, $c(\alpha)$.

     For  Neumann boundary conditions we failed to find in the
literature the formula analogous to equation\ (\ref{eq4-7}), i.e.,
the asymptotic expansion of the heat kernel for the operator
$-\Delta $ defined in the region of a plane with smooth boundary
curve. However, as was noted earlier, the spectral problem on the
plane, envisaged by us, and its cylindrical generalization have the
same coefficients $B_n$.  Therefore, for verification of our
results, concerning  the contributions to $B_n$ due to the smooth
parts of the boundary, we used the expansion of the heat kernel
for Robin conditions with smooth boundary that has the dimension
greater than 1 (see references~\cite{ER,KCD}).

\section{Heat kernel coefficients for external sector}
\subsection{Basic formulas}

Here we shall use the technique
applied in \cite{Moss}. It is close to the
method developed in  \cite{Kirsten,Bordag,BKD} and employed in
preceding subsections.

  Let us consider the spectral zeta function depending on a parameter $x^2$
\begin{equation}
\label{eq3-47}
\zeta(s,x^2)=\sum_{n}(\lambda_n^2+x^2)^{-s}{.}
\end{equation}
It may be regarded as an extension to the general spectral problem
of the Epstein-Hurwitz zeta function
\[
\zeta_{{\rm EH}}(s,a^2)=\sum_{n=1}^{\infty}(n^2+a^2)^{-s}{.}
\]

It turns out that the heat kernel coefficients $B_n$ can be found from the
expansion of the  function $\zeta(s,x^2)$ in terms of inverse powers of $x$
developed for a certain value of $s$. It is convenient to chose this value
to be equal to
$d/2$. In fact, from the definition of the  gamma function it follows that
\begin{equation}
\label{eq3-48}
\Gamma (s)\,(\lambda^2+x^2)^{-s}=\int_{0}^{\infty}\!\!rmd t\,t^{s-1}
e^{-(\lambda^2+x^2)t}{.}
\end{equation}
For $s=1+d/2$ equation\  (\ref{eq3-48}) gives
\begin{equation}
\label{eq3-49} \fl \Gamma\left( 1+\frac{d}{2} \right )
\sum_{n}(\lambda_n^2+x^2)^{-1-d/2}=\int_{0}^{\infty}\!\!\rmd t\,t^{d/2}e^{-x^2t}
\sum_{n}e^{-\lambda_n^2t}=\int_{0}^{\infty}\!\!\rmd t\,t^{d/2}e^{-x^2t}K(t)\,{.}
\end{equation}
On substituting  the asymptotic expansion (\ref{eq2-5}) in
equation\ (\ref{eq3-49}) we obtain
\begin{eqnarray}
\fl \Gamma( 1+{d}/{2} )\zeta ( 1+{d}/{2},x^2 )
&\simeq&\sum_{n=0}^{\infty}\frac{B_{n/2}}{(4\pi)^{d/2}} \Gamma
\left ( 1+\frac{n}{2}
\right)      x^{-n-2} \nonumber\\
\fl &=&\frac{1}{(4\pi)^{d/2}}\left [
\frac{B_0}{x^2}+\frac{B_{1/2}\Gamma(3/2)}{x^3}+\frac{B_1\Gamma(2)}{x^4}
+\frac{B_{3/2}\Gamma(5/2)}{x^5}\nonumber  \right .\\
\fl && \left .
+\frac{B_2\Gamma(3)}{x^6}+\frac{B_{5/2}\Gamma(7/2)}{x^7}+\frac{B_3\Gamma(4)}{x^8}
+{\cal O} (x^{-9}) \right ]{.} \label{eq3-50}
\end{eqnarray}

Let $  F(z)=0$ be the frequency equation which determines the
spectrum $\lambda_n$ in the problem under consideration. We also
suppose that the function $F(z)$ allows one to rewrite this
equation in the form
\begin{equation}
\label{eq3-52} \prod_{n}(\lambda_n^2-z^2)=0\,{.}
\end{equation}
Taking into account the equality
\begin{equation}
\label{eq3-53}
\frac{1}{(\lambda_n^2+x^2)^m}=-\frac{(-1)^m}{\Gamma(m)}\left(
\frac{\rmd}{2x\, \rmd x} \right )^m    \ln (\lambda_n^2+x^2)\,{,}\quad
z=ix\,{,}
\end{equation}
we recast the left-hand side of equation\ (\ref{eq3-50})  to the
form
\begin{equation}
\label{eq3-54}
\Gamma
\left (1+\frac{d}{2}
\right )\zeta
\left (       1+\frac{d}{2}, x^2
\right )=-\left (-\frac{1}{2x}\frac{\rmd}{\rmd x}
\right )^{1+d/2}\ln F(ix)\,{.}
\end{equation}
Obviously formula (\ref{eq3-54}) is applicable only to the manifolds of even
dimension.

 Rather than to calculate the heat kernel coefficients
$B_n^{II}$ for the external sector alone it is simpler to find
first the coefficients $B_n^{I+II}$ for the union of the sectors I
and II. Then the coefficients $B_n^{II}$ are obtained as the
corresponding differences
\[
B_n^{II}=B_n^{I+II} -B_n^{I}, \quad n=0,\,1/2,\, 1,\,\ldots \,{.}
\]

When calculating the heat kernel coefficients for the union of the
internal (I) and external (II) sectors on should take into account
the following. The technique used by
us~\cite{Kirsten,Bordag,BKD,Moss,Esposito} gives the
difference between the zeta function  for the region I+II and the
zeta function for the corresponding part of the Euclidean space.
The last contribution  is usually referred to as the Minkowski
space-time contribution~\cite{BEKL}. In the case under
consideration the zeta function for an open angle $\alpha$
(without circular arc 1-2) is subtracted from the zeta function
sought for. The heat kernel coefficients $\bar B_n$, corresponding
to the Minkowski space-time contribution,  can be calculated in
the following way: in the respective heat kernel coefficients for
the sector I one should put $R=R_1\to\infty$ and discard the curvature
contribution of the arc 1-2 and of two rightangled internal
corners at the points 1 and 2.

\subsection{Internal and external circular sectors of a wedge with Dirichlet condition
on separating arc}

   Now we proceed to practical using the general formulas
(\ref{eq3-50}) and (\ref{eq3-54}) for calculation of the heat
kernel coefficients.  We consider the scalar Laplace operator on
the union of internal and external circular sectors on a plane
(see figure\ 1) with Dirichlet conditions on the arc 1--2
separating these sectors. As it was explained in Section 2 the
heat kernel coefficients for the corresponding boundary value
problem in space ($d=3$)  The frequency equations are will be the
same.
\begin{eqnarray}
J_{np}(Rz)=0,& (\mbox{internal sector I}), \label{eq3-55} \\
H^{(1)}_{np}(Rz)=0, & (\mbox{external sector II}), \label{eq3-56}
\quad
 n=1,\,2,\, \ldots \,{,} \quad p=\pi/\alpha{.}
\end{eqnarray}

     Further we shall concern with the product
$J_\nu(ix)\,H^{(1)}_\nu(ix)$ and use for it the uniform asymptotic
expansion~\cite{AS} which depends only on $x^2$. In this case the
condition (\ref{eq3-52}) is apparently satisfied. This point can
also be explained in the following way. In view of the
formula\cite{GR,Watson}
\begin{equation}
\label{eq3-57} J_\nu(z)=\frac{(z/2)^\nu}
{\Gamma(\nu+1)}\prod_{m=1}^{\infty} \left(
1-\frac{z^2}{z^2_{\nu,m}} \right ){,}\quad \nu\not= -1,\,-2,\,
\ldots
 \end{equation}
for the function $z^{-\nu}J_\nu (z)$ the representation of type
(\ref{eq3-52}) holds. Here $z_{\nu,m}\; (m=1,\,2,\,3,\,\ldots)$
are the nonzero roots of the function $J_\nu(z)$. The multiplier
$(z/2)^\nu$ in equation\ (\ref{eq3-57}) is canceled in the product
$J_\nu (ix) H^{(1)}_\nu(ix)\sim I_\nu(x)K_\nu(x)$ with the
multiplier
 $(x/2)^{-\nu}$ following from the small $x$ asymptotics of the
function $K_\nu(x)$: $K_\nu(x)\sim (1/2)\Gamma (\nu)(x/2)^\nu,\; \nu>0$.
Hence the requirement (\ref{eq3-52}) is satisfied.

 Setting in (\ref{eq3-54}) $d=2$ and substituting in it
the left-hand sides of the frequency equations (\ref{eq3-55}) and
(\ref{eq3-56}) we obtain
\begin{equation}
\label{eq3-58}
\zeta_D(2,x^2)=-
\left (  \frac{1}{2x}\frac{\rmd}{\rmd x}
\right )^2 \sum_{n=1}^{\infty}\ln I_\nu(Rx)K_\nu(Rx)\,{.}
\end{equation}
Now we use the uniform asymptotic expansion for the product of the
modified Bessel functions~\cite{AS}
\begin{equation}
\label{eq3-59}
\ln I_\nu (Rx) K_\nu(Rx)=-\ln 2\nu +\ln t+\sum_{j=1}^{\infty}
\frac{G^D_{2j}(t)}{\nu^{2j}}{,}
\end{equation}
where
\begin{equation}
\label{eq3-60}
t=\frac{1}{\sqrt{1+z^2}},\quad z=\frac{Rx}
{\nu}{,}\quad \nu=np=n\frac{\pi}{\alpha}{,}
\end{equation}
and $G_j^D(t)$ are the  polynomials in $t$ expressed in terms of
the known functions $u_k(t)$. In order to calculate the first six
coefficients in the expansion (\ref{eq3-50}) it is sufficient to
keep two terms of the sum in (\ref{eq3-59}). The relevant
coefficients $G_2^D(t)$ and  $G_4^D(t)$ are
\begin{eqnarray}
G^D_2(t)&=&\frac{1}{8}t^2 -\frac{3}{4} t^4+\frac{5}{8}t^6{,} \\
G^D_4(t)&=&\frac{13}{64}t^4- \frac{71}{16}t^6+\frac{531}{32}t^8-
\frac{339}{16}t^{10}
+\frac{565}{64}t^{12}\,{.}
\end{eqnarray}
Substituting in  (\ref{eq3-58})  differentiation with respect to
$x$ by differentiation with respect to $t$ we get
\begin{eqnarray}
\zeta_D(2,x^2)&=&-\sum_{n=1}^{\infty}\left (\frac{R}{\nu}
\right )^4\left (\frac{t^3}{2}\frac{\rmd}{\rmd t}
\right )^2
\left [\ln t+\sum_{j=1}^2{}\frac{G_{2j}^D(t)}{\nu^{2j}}
\right ]                                             \nonumber \\
&=&-R^4\sum_{n=1}^\infty{}\left [\frac{t^4}{2\nu^4}+\frac{t^6}{4\nu^6}(1-18t^2
+30t^4)
\right . \nonumber \\
&&\left .+\frac{3t^8}{32\nu^8}(13-568t^2+3540t^4-6780t^6+3955t^8)
\right ]{.}     \label{eq3-63}
\end{eqnarray}
It will be recalled that $t$ depends on $x$ and $n$ through
(\ref{eq3-60}).

    All the sums in the equation\ (\ref{eq3-63}) are finite. Hence the problem
of analytic continuation does not emerge here. In order to do the
summation exactly we use the formula~\cite{GR,BY}
\begin{equation}
\label{eq3-64}
\sum_{n=1}^{\infty}\frac{1}{y^2+n^2}=\frac{\pi}{2y}\left ( \coth \pi
y-\frac{1}{\pi y}
 \right )  \simeq \frac{\pi }{2y}-\frac{1}{2y^2}\equiv
 S_1(y)\,{.}
\end{equation}
     When $y\to \infty$, the function $S_1(y)$ affords the value of
the sum on the left-hand side of equation\ (\ref{eq3-64}) up to
exponentially small corrections. Step-by-step differentiation with
respect of $y$ of the left and right hand sides of the equation\
(\ref{eq3-64}) gives the values of all the sums entering
(\ref{eq3-63})
\begin{eqnarray}
\sum_{n=1}^{\infty}\frac{1}{(y^2+n^2)^2}&\simeq&-\frac{1}{2y}\,
\frac{\rmd}{\rmd y} S_1(y)=\frac{\pi}{4y^3}-\frac{1}{2y^4}\equiv S_2(y)\,{,}
\nonumber \\
\sum_{n=1}^{\infty}\frac{1}{(y^2+n^2)^3}&\simeq&-\frac{1}{2}\,\frac{1}{2y}\,
\frac{\rmd}{\rmd y} S_2(y)=\frac{3\pi}{16y^5}-\frac{1}{2y^6}\equiv S_3(y)\,{,}
\nonumber \\
\nonumber \\
\cdots\cdots && \cdots\cdots\cdots\cdots \cdots\nonumber \\
\nonumber \\
\sum_{n=1}^{\infty}\frac{1}{(y^2+n^2)^{k+1}}&\simeq&-\frac{1}{k}\,\frac{1}{2y}\,
\frac{\rmd}{\rmd y} S_k(y)\equiv S_{k+1}(y)\,{.}\label{eq3-65}
\end{eqnarray}

     In order to express the zeta function $\zeta_D(2,x^2)$ in terms of
$S_k(x),\; k=1,\,2,\,\ldots \,{,} \,8 $ explicitly the following
substitutions should be accomplished in equation\ (\ref{eq3-63})
\begin{equation}
\label{eq3-66} \frac{t^2}{\nu ^2}=\frac{1}{p^2\,(n^2+{\tilde
x}^2)}\,{,}
\end{equation}
\begin{equation}
\label{eq3-67} t^2=1-\frac{{\tilde x}^2}{n^2+{\tilde x}^2}\,{,}
\quad {\tilde x}=\frac{Rx}{p}\,{,}
\end{equation}
the change (\ref{eq3-67}) being done only in the round brackets in
this equation. As a result the zeta function $\zeta _D(2,x^2)$ acquires
the form
\begin{eqnarray}
\fl \zeta_D(2,x^2)&=&- \frac{R^4}{2p^4}S_2({\tilde
x})-\frac{R^4}{4p^6}\left [13\, S_3({\tilde x})-42 \,{\tilde x}
^2\,S_4({\tilde x})+30\,{\tilde x}^4\,S_5({\tilde x})
\right ]                     \nonumber \\
\fl &&-\frac{3R^4}{32p^8}\left [ 160\, S_4({\tilde x})
-1992\,{\tilde x}^2\,S_5({\tilde x})+6930\,
 {\tilde x}^4\,S_6({\tilde x}) \right . \nonumber \\
\fl &&\left .-9040\,{\tilde x}^6\,S_7({\tilde x})+3955\,{\tilde
x}^8\,S_8({\tilde x}) \right ]{.} \label{eq3-68}
\end{eqnarray}
     Substitution of the explicit expressions
for the functions $S_k({\tilde x})$ from (\ref{eq3-65}) to
(\ref{eq3-68})   gives
\begin{eqnarray}
\fl \zeta_D(2,x^2)&=&-\frac{\alpha
R}{8x^3}+\frac{1}{4x^4}+\frac{3\,\alpha}{512Rx^5}
+\frac{1}{8R^2x^6}+\frac{555\,\alpha}{2^{17}R^3x^7}+\frac{39}{2^6R^4x^8}+
{\cal O}(x^9)\,{.}\label{eq3-69}
\end{eqnarray}
Comparing the expansions (\ref{eq3-50}) and (\ref{eq3-69}) we obtain
the values of the first seven heat kernel coefficients  in the
problem under consideration
\[
\fl
\bar B_0^{I+II}=0,\quad \bar B^{I+II}_{1/2}=-\alpha R\sqrt \pi,\quad
\bar B_1^{I+II}=\pi,
\]
\begin{equation}
\label{eq3-70} \fl
\bar B_{3/2}^{I+II}= \frac{\alpha \sqrt \pi}{32\,R}, \quad
 \bar B_2^{I+II}=\frac{\pi}{4R^2},\quad \bar B^{I+II}_{5/2}
=\frac{37\,\sqrt \pi\,\alpha}{2^{12}R^3},
\quad \bar B_3^{I+II}=\frac{13\pi}{32R^4}\,{,}
\end{equation}
where $\bar B_n^{I+II}=B_n^{I+II}-\bar B_n$, and $\bar B_n$ are
the heat kernel coefficients corresponding to
the open angle $\alpha $  (without the arc 1-2)
\begin{equation}
\label{eq5-24}
\fl
\bar B_{0}=\frac{1}{2}\,\alpha R^2_1,\quad
\bar B_{1/2}=-\sqrt \pi R_1 -\frac{\sqrt \pi}{2}\,\alpha R_1,\quad
\bar B_{1}=\frac{\pi^2-\alpha ^2}{6\alpha}=c(\alpha),\quad R_1\to \infty\,{.}
\end{equation}
All the rest of coefficients $\bar B_j$ vanish $\bar B_j=0,\quad j\ge3/2$.

Using equations (\ref{eq3-70}), (\ref{eq5-24}) and the results for the
internal sector I, presented in  table 1 we deduce
the heat
kernel coefficients for the external sector II with Dirichlet boundary
condition
\begin{eqnarray}
\label{eq5-25}
\fl B_0^{II}=\frac{\alpha}{2}\,(R_1^2- R^2), \quad B_{1/2}^{II}=-\sqrt \pi\,(R_1-R) -
\frac{\alpha \sqrt \pi}{2}\,(R_1+R), \quad
B_1^{II}=\frac{\pi}{2}-\frac{\alpha}{3}\,{,}
\nonumber \\
\label{eq3-71} \fl  B^{II}_{3/2}= -\frac{\sqrt
\pi}{4\,R}+\frac{\alpha\,\sqrt \pi}{64\,R}, \quad
 B^{II}_2=\frac{\pi}{8\,R^2}-\frac{4}{315}\,\frac{\alpha}{R^2}, \quad B^{II}_{5/2}
=-\frac{25}{96}\,\frac{\sqrt \pi}{R^3}+\frac{37\,\alpha\,\sqrt \pi}{2^{13}R^3}\,{.}
\end{eqnarray}

\subsection{Internal and external circular sectors
of a wedge with Neumann condition
on separating arc}
In the case of Neumann boundary conditions on a separating arc the
heat kernel coefficients for the union of internal and external
sectors are calculated completely in a similar line. Now the
frequency equations read
 \begin{eqnarray}
\fl \left.\frac{\rmd}{\rmd r}J_{np}(rz)\right |_{r=R}&=&0,\quad
(\mbox{internal sector I}), \label{eq3-72} \\
\fl  \left.\frac{\rmd}{\rmd r}H^{(1)}_{np}(rz)\right |_{r=R}&=&0, \quad
(\mbox{external sector II}), \label{eq3-73} \quad n=0,\, 1,\,2,\,
\ldots \,{,} \quad p=\pi/\alpha .
\end{eqnarray}
The angular part of the eigenfunctions is proportional to $\cos (np\theta)$,
therefore the index $n$ takes integer values starting with zero.

For the zeta function in this eigenvalue problem  the
representation analogous to (\ref{eq3-58}) holds
\begin{equation}
\label{eq3-74}
\zeta_N(2,x^2)=-\left (
\frac{1}{2x}\,\frac{\rmd}{\rmd x}
\right )^2               \sum_{n=0}^{\infty}\ln
\left [ -I'_{np}(Rx)K'_{np}(Rx)
\right ]{.}
\end{equation}
Again we use the uniform asymptotic expansion \cite{AS}
\begin{equation}
\label{eq3-75}
\ln \left [ -I'_{np}(Rx)K'_{np}(Rx)
\right ] \simeq -\ln 2 \nu - \ln t+
\sum_{j=1}^{2}\frac{G^N_{2j}(t)}{\nu^{2j}}\,{,}
\end{equation}
where
\begin{eqnarray}
G_2^N(t)&=&-\frac{3}{8}t^2+\frac{5}{4}t^4-\frac{7}{8}t^6{,}
\nonumber \\
G_4^N(t)&=&-\frac{27}{64}t^4+\frac{109}{16}t^6-\frac{733}{32}t^8+\frac{441}{16}
t^{10}-\frac{707}{64}t^{12}{.}\label{eq3-76}
\end{eqnarray}
On substituting  (\ref{eq3-75}) and (\ref{eq3-76}) and
differentiating in equation\ (\ref{eq3-74}),  the zeta function
under consideration assumes the form
\begin{eqnarray}
\fl \zeta_N(2,x^2)&=&-R^4\sum_{n=0}^{\infty}\left [
-\frac{t^4}{2\nu ^4}-\frac{3t^6}{4\nu^6}\,(1-10\,t^2+14\,t^2)
\right . \nonumber \\
\fl &&\left . -\frac{t^8}{32\nu^8}\,(81-2616\,
t^2+14660\,t^4-26460\, t^6+14847\, t^8) \right ]{.}\label{eq3-77}
\end{eqnarray}
In order to take the sum over $n$ in  (\ref{eq3-77}) we apply the
formula that follows from  (\ref{eq3-64})
\begin{equation}
\label{eq3-78}
\sum_{n=0}^{\infty}\frac{1}{y^2+n^2}=\frac{\pi}{2y}\left (
\coth \pi y+\frac{1}{\pi y}
\right )\simeq \frac{\pi}{2y} +\frac{1}{2y^2}\equiv \bar S_1(y)\,{.}
\end{equation}
Sequential differentiation of equation\ (\ref{eq3-78}) gives
\begin{equation}
\label{eq3-79} \sum_{n=0}^{\infty}\frac{1}{(y^2+n^2)^{k+1}}\simeq-
\frac{1}{k} \, \frac{1}{2y}\,\frac{\rmd}{\rmd y}\bar S_k(y)\equiv \bar
S_{k+1}(y)\,{.}
\end{equation}
On making use of the change of variables (\ref{eq3-66}) and
(\ref{eq3-67}) in equation\ (\ref{eq3-77}) the zeta function
$\zeta_N(2,x^2)$ assumes the form
\begin{eqnarray}
\zeta_N(2,x^2)&=&\frac{R^4}{2p^4}\bar S_2({\tilde
x})+\frac{3R^4}{4p^6} \left [ 5\bar S_3({\tilde x})-18\,{\tilde
x}^2\,\bar S_4({\tilde x})+14\,{\tilde x}^4\,\bar S_5({\tilde x})
\right ]                      \nonumber \\
&&+\frac{R^4}{32p^8}\left [512\,\bar S_4({\tilde x})-6712\,{\tilde
x}^2\,\bar S_5({\tilde x})+ 24362\, {\tilde x}^4\,\bar S_6({\tilde
x})
\right . \nonumber \\
&&\left .-32928\,{\tilde x}^6\,\bar S_7({\tilde x})+14847\,{\tilde
x}^8\,\bar S_8({\tilde x})\right ]{.} \label{eq3-80}
\end{eqnarray}
Substitution of the function $\bar S_k({\tilde x})$ with
$k=2,\,3,\,\ldots\,, 8$ from  (\ref{eq3-78}) and (\ref{eq3-79})
gives
\begin{equation}
\label{eq3-81} \fl
\zeta_N(2,x^2)=\frac{\alpha\,R}{8\,x^3}+\frac{1}{4\,x^4}+\frac{15}{512}\,
\frac{\alpha}{R^2\,x^5}+\frac{3}{8}\,\frac{1}{R^2\,x^6}+
\frac{4035}{131072}\,\frac{\alpha}{R^3\,x^7}+
\frac{81}{64}\,\frac{1}{R^4\,x^8}+{\cal O}(x^{-9})\,{.}
\end{equation}
Comparison of the expansions (\ref{eq3-81}) and (\ref{eq3-50})
 gives the following values for
the heat kernel coefficients $\bar B_n^{I+II}$ for Neumann boundary conditions
\[\fl
\bar B^{I+II}_0=0,\quad \bar B^{I+II}_{1/2}=\alpha R\sqrt \pi,\quad
\bar B^{I+II}_1=\pi,
\]
\begin{equation}
\label{eq3-82}
\fl
\bar B^{I+II}_{3/2}= \frac{5\,\alpha \,\sqrt \pi}{32\,R}, \quad
\bar B^{I+II}_2=\frac{3\,\pi}{4R^2},\quad \bar B^{I+II}_{5/2}
=\frac{269}{4096}\,\frac{\alpha\,\sqrt \pi}{R^3}{,}
\quad \bar B^{I+II}_3=\frac{27\,\pi}{32\,R^4}\,{.}
\end{equation}

The heat kernel coefficients corresponding to the Minkowski
space-time contribution in the case under consideration are
derived from the respective  coefficients for internal sector I
by putting there $R=R_1\to\infty$ and omitting
the contribution due to the curvature of the arc 1-2 and
contributions of the rightangled corners at the points 1 and 2 (see Tables  2 and 3)
\begin{equation}
\label{eq5-37}
\fl
\bar B_0=\frac{\alpha}{2}\,R^2,\quad
\bar B_{1/2}=\sqrt \pi R_1+\frac{\alpha}{2}\,\sqrt \pi R_1 \quad
\bar B_1=c(\alpha),\quad
\bar B_j=0,\quad j=3/2,\,2,\dots \,{.}
\end{equation}

By making use of the equations (\ref{eq3-82}), (\ref{eq5-37}) and  table 2 we derive
the heat kernel
coefficients for the external sector II with Neumann condition
\[ \fl
B^{II}_0=\frac{\alpha}{2}\,(R_1^2- R^2),\quad B^{II}_{1/2}=\sqrt \pi\,(R_1- R)+
 \frac{\alpha}{2}\,\sqrt \pi\,(R_1+R)\quad B^{II}_1=\frac{\pi}{2}-
\frac{\alpha}{3}\,{,}
\]
\begin{equation}
\label{eq3-83} \fl B^{II}_{3/2}= -\frac{3\,\sqrt
\pi}{4\,R}+\frac{5\,\alpha\,\sqrt \pi}{64\,R}, \quad
 B^{II}_2=\frac{3\,\pi}{8\,R^2}-\frac{4}{45}\,\frac{\alpha}{R^2},\quad B^{II}_{5/2}
=-\frac{21}{32}\,\frac{\sqrt \pi}{R^3}+\frac{269\,\alpha\,\sqrt \pi}{2^{13}R^3}\,{.}
\end{equation}

\subsection{Internal and external sectors on a cone}
In the case of a cone the eigenfunctions are defined in
(\ref{eq3-34}) with $\lambda_{nm}$ being the roots of the
frequency equations (\ref{eq3-55}), (\ref{eq3-56}) and
(\ref{eq3-72}),  (\ref{eq3-73}) where $p=2\,\pi/\alpha$. For both
Dirichlet and Neumann conditions on the circle 1--2 the index $n$
ranges from $n=0$. The state degeneracy in both the cases
 is determined by equation\ (\ref{eq3-37}).

All this implies that in order to proceed to a cone one should put
in the relevant formulas for a wedge $p=2\,\pi/\alpha$ and  use
for summation over $n$ the following relations (instead of
equations\ (\ref{eq3-65}),
 (\ref{eq3-78}), and (\ref{eq3-79}))
\begin{eqnarray}
\fl
&&\frac{1}{y^2}+2\sum_{n=1}^{\infty}\frac{1}{y^2+n^2}=\frac{\pi}{y}\coth
\pi\,y
\simeq \frac{\pi}{y}\equiv S_1^c(y)\,{,} \nonumber \\
\fl&&
\frac{1}{y^{2(k+1)}}+2\sum_{n=1}^{\infty}\frac{1}{(y^2+n^2)^{k+1}}\simeq
-\frac{1}{k}\,\frac{1}{2y}\,\frac{\rmd}{\rmd y}S_k^c(y)\equiv
S^c_{k+1}(y),\quad k=1,\,2,\, \ldots\,{.} \label{eq3-84}
\end{eqnarray}
   It is essential that these summation formulas should be employed
for both  Dirichlet and Neumann boundary conditions given on the circle 1--2
and separating the internal and external sectors of a cone.

     First we consider  Dirichlet conditions. Carrying out in
(\ref{eq3-68}) the change of variables (\ref{eq3-38}) and using
for summation the functions $S_k^c({\tilde x})$ from equation\
(\ref{eq3-84}) we obtain the  expansion for the zeta function in
the spectral problem at hand
\begin{equation}
\label{eq3-85}
\zeta_D^c(2,x^2)=-\frac{\alpha\,R}{8\,x^3}+\frac{3\,\alpha}{512\,R\,x^5}+
\frac{555}{2^{17}}\,\frac{\alpha}{R^3\,x^7}+{\cal O}(x^{-9})\,{.}
\end{equation}
This expansion can formally be derived from the relevant zeta
function for a wedge, equation\ (\ref{eq3-69}), by omitting the
terms with even powers of $x$. Comparison of the series
(\ref{eq3-85}) with equation\ (\ref{eq3-50}) affords the values of
the first seven heat kernel coefficients $\bar B_n^{I+II}$
\begin{eqnarray}
\label{eq3-86}
\bar B^{I+II}_0=\bar B^{I+II}_1=\bar B^{I+II}_2=\bar B^{I+II}_3=0,\;\;
\bar B^{I+II}_{1/2}=-\alpha\,\sqrt
\pi R, \nonumber \\
\bar  B^{I+II}_{3/2}= \frac{\alpha\,\sqrt \pi}{32R}, \;\;
\bar   B^{I+II}_{5/2}
=\frac{37\,\alpha\,\sqrt \pi}{2^{12}R^3}\,{.}
\end{eqnarray}
The heat kernel coefficients for Minkowski spacetime contribution are
\begin{equation}
\label{eq5-42}
\fl
\bar B_0=\frac{\alpha}{2}\,R_1^2,\quad
\bar B_{1/2}=-\frac{\alpha}{2}\,\sqrt \pi \,R_1,\quad
\bar B_1=2\,c(\alpha/2),\quad \bar B_j=0,\;\; j\ge3/2\,{.}
\end{equation}

We deduce from equations  (\ref{eq3-86}), (\ref{eq5-42}),
and  (\ref{eq3-45a})
the heat kernel  coefficients for external sector of a cone
with Dirichlet boundary conditions on the circle 1--2
\[
B^{II}_0=\frac{\alpha}{2}\,(R_1^2-R^2),\quad
B^{II}_{1/2}=-\frac{\alpha}{2}\,\sqrt \pi\,(R_1-R),\quad
B^{II}_1=-\frac{\alpha}{3}{,}
\]
\begin{equation}
\label{eq3-87}
B^{II}_{3/2}= \frac{\alpha\,\sqrt \pi}{64\,R}, \quad B^{II}_2=
-\frac{4\,\alpha}{315\,R^2},\quad
 B^{II}_{5/2} =\frac{37\,\alpha\,\sqrt \pi}{2^{13}R^3}{.}
\end{equation}

By analogy with this we obtain  from equation\ (\ref{eq3-80})
for  Neumann boundary conditions
\begin{equation}
\label{eq3-88}
\zeta_N^c(2,x^2)=\frac{\alpha\,R}{8\,x^3}+\frac{15\,\alpha}{512\,R\,x^5}+
\frac{4035}{2^{17}}\,\frac{\alpha}{R^3\,x^7}+{\cal O}(x^{-9})\,{.}
\end{equation}
This expansion can formally be derived from the relevant zeta
function for a wedge equation\ (\ref{eq3-81}) by omitting the
terms with even powers of $x$. Comparing the equations\
(\ref{eq3-88}) and (\ref{eq3-50}) we obtain  the values of the
first seven heat kernel coefficients $\bar B_n^{I+II}$
\begin{eqnarray}
 \label{eq3-89} \bar B^{I+II}_0=
\bar B^{I+II}_1=
\bar B^{I+II}_2=\bar B^{I+II}_3=0,\;\;
\bar  B^{I+II}_{1/2}=\alpha\,\sqrt
\pi R,\nonumber \\
\bar  B^{I+II}_{3/2}= \frac{5\,\alpha\,\sqrt \pi}{32R}, \;\;
\bar   B^{I+II}_{5/2}
=\frac{269\,\alpha\,\sqrt \pi}{2^{12}R^3}\,{.}
\end{eqnarray}
The heat kernel coefficients for Minkowski spacetime contribution in this case are
\begin{equation}
\label{eq5-46}
\fl
\bar B_0=\frac{\alpha}{2}\,R_1^2,\quad
\bar B_{1/2}=\frac{\alpha}{2}\,\sqrt \pi \,R_1,\quad
\bar B_1=2\,c(\alpha/2),\quad \bar B_j=0,\;\; j\ge3/2\,{.}
\end{equation}

From equations (\ref{eq3-89}), (\ref{eq5-46}), and (\ref{eq3-46a})
we derive the heat kernel  coefficients for the external sector II of the
cone with Neumann boundary conditions on the circle 1--2
\[
B^{II}_0=\frac{\alpha}{2}\,(R_1^2-R^2),\quad
B^{II}_{1/2}=\frac{\alpha}{2}\,\sqrt \pi \,(R_1-R),\quad
B^{II}_1=-\frac{\alpha}{3}{,}
\]
\begin{equation}
\label{eq3-90}
B^{II}_{3/2}= \frac{5\,\alpha\,\sqrt \pi}{64\,R}, \quad B^{II}_2=
\frac{4\,\alpha}{45\,R^2},
\quad
 B^{II}_{5/2} =\frac{269\,\alpha\,\sqrt \pi}{2^{13}R^3}\,{.}
\end{equation}

\section{Specification of individual contributions to heat kernel
coefficients for external sector. Some general rules}

   The heat kernel coefficients for external sector are given in
equations (\ref{eq5-25}), (\ref{eq3-83}), (\ref{eq3-87}), and
(\ref{eq3-90}). Identification of the individual contributions to
these coefficients due to the boundary nonsmoothness does not
differ basically from the analogous procedure for internal sector
carried out in section 4. The sole point that should be noted here
is the following. The area $|\Omega|$ and the perimeter $L$ are in
this case infinite. Therefore they should be treated as the limit,
when   $R_1$ tends to infinity, of the  expressions
\begin{equation}
\label{eq6-1} |\Omega|=\frac{\alpha}{2}\,(R_1^2-R^2),\quad
L=2\,(R_1-R)+\alpha (R_1+R)\,{.}
\end{equation}
The coefficients $B_0$ and $B_{1/2}$ are defined, as before, by
equations (\ref{eq4-a}), (\ref{eq4-b}), and (\ref{eq4-13a}) with
$|\Omega|$ and $L$ defined in (\ref{eq6-1}). Contributions to the
rest of the heat kerenel coefficients, which are proportional to
the angle $\alpha$ are due to the curvature of the arc 1--2 and
the contributions which are independent of $\alpha$ are due to the
rght-angled corners at the points 1 and 2. The results  of this
analysis are represented in table 3 together with the heat kernel
coefficients for the internal sector I. Analysis of this table
enables one to reveal some general rules for corner contributions.

     It is interesting to note that the right-angled corners at
the points 1 and 2 give the contributions to the coefficients
$B_{3/2}$ and $B_{5/2}$ which have opposite signs for internal and
external sectors, it being valid for all configurations and
boundary conditions considered. Such a behaviour  seems to be
related with convexity (internal sector) or concavity  (external
sector) of the arc 1--2.  It is analogous to the contributions of
the smooth segments of the boundary to the heat kernel
coefficients $B_1$ and $B_2$.

     The boundary discontinuities due to the corner at the origin
$O$ and to the corners at the points 1 and 2 contribute to the heat
kernel coefficients basically in different ways. Corner at the origin
gives contribution only to the coefficient $B_1$ (even when
$\alpha=\pi/2$). The corners at the points 1 and 2 contribute to all
the heat kernel coefficients starting with $n=1$. Obviously the
reason of this distinction is a geometrical one, the corner at the
origin is formed by crossing two straight lines, while the corners at
the points 1 and 2 are the result of intersection of lines one of
which has a  nonzero curvature.

This general assertion concerning the corner contribution to the heat
kernel coefficients can be illustrated by a known expression for the
heat kernel expansion for a rectangle with sides $a$ and $b$
\begin{eqnarray}
\label{eq4-14} \fl
K(t)&=&\sum_{m=1}^{\infty}\sum_{n=1}^{\infty}\exp\left [- \left
(\frac{m^2}{a^2}+\frac{n^2}{b^2} \right )\pi^2t \right ] = \left
(\frac{a}{\sqrt{4\pi t}}-\frac{1}{2} \right ) \left
(\frac{b}{\sqrt{4\pi t}}-\frac{1}{2}
\right ) +\mbox{ES}  \nonumber \\
\fl &=& \frac{ab}{4\pi t} -\frac{a+b}{4\sqrt{\pi t}}+\frac{1}{4}+
\mbox{ES}\,{,}
\end{eqnarray}
where ES denotes the exponentially small corrections as $t\to 0$
(see, for example, \cite{W}). Here the scalar operator $-\Delta$
with Dirichlet boundary conditions is considered. For a rectangle
the coefficient $B_1$ is obviously equal to the contributions of
four right-angled corners. Indeed, the third term in the expansion
(\ref{eq4-14}) can be represented in the form
\[
\frac{1}{4}=\frac{1}{4\,\pi}\,B_1=\frac{1}{4\,\pi}\,4
\,c(\alpha=\pi/2)\,{,}
\]
where $c(\alpha)$ is given in equation\ (\ref{eq4-5}). Besides
these corners the boundary of a rectangle has no other
singularities, therefore the heat kernel coefficients $B_n$ with
$n>1$ vanish in this problem.

These rules for obtaining the heat kernel coefficients are
directly generalized to an arbitrary polygon with the angles
$\alpha_i$. The first two coefficients $B_0$ and $B_{1/2}$ are
defined by equations\ (\ref{eq4-a}) and (\ref{eq4-b}),
respectively, where $|\Omega|$ is the area of the polygon and $L$
is its perimeter. The third coefficient $B_1$ is equal to the sum
of the contributions due to the angles   $\alpha_i$
\begin{equation}
\label{eq4-16a}
B_1=\sum_{i}c(\alpha_i)\,{.}
\end{equation}
The rest of the coefficients $B_n,\;n>1$ vanish. In particular, it implies
that the zeta function technique should  provide  a finite value of
the Casimir energy for a polygon on a plane ($B_{3/2}=0$) and for a
cylindrical generalization of the polygon spectral problem ($B_2=0$).
These subjects have been discussed earlier in papers~\cite{Dow1}.

    In reference\ \cite{NLS-1} the vacuum energy of massless fields
including electromagnetic field was  calculated for the boundary
configuration shown in figure~1 with $\alpha=\pi$. Both versions
of this boundary value problem were considered, three-dimensional
one (a semi-circular infinite cylinder) and two-dimensional
spectral problem on the plane. In both the cases the zeta
regularization didn't give a finite value of the Casimir energy.
As known~\cite{Kirsten}, the reason of this  is nonzero heat
kernel coefficients $B_2$ for $d=3$ and $B_{3/2}$ for $d=2$. Using
the results of the present paper we can elucidate the geometrical
origin of this fact. Let us consider electromagnetic filed in
internal and external sectors of the wedge with $\alpha =\pi$. For
the boundaries with cylindrical symmetry electromagnetic field
reduces to  two massless fields subjected to Dirichlet and Neumann
boundary conditions~\cite{HdP1,Str}. From table 3 it follows that
\begin{equation}
\label{eq4-15} \fl
B_2^{\mbox{e-m}}=2\,\frac{\pi}{8\,R^2}+2\,\frac{3}{8}\,\frac{\pi}{R^2}
=\frac{\pi}{R^2}\,{,}\quad  B_{3/2}^{\mbox{e-m}}=2\frac{\pi\sqrt
\pi}{64\,R} +2\,\frac{5}{64}\,\frac{\pi \sqrt
\pi}{R}=\frac{3}{16}\, \frac{\pi \sqrt \pi}{R}\,{.}
\end{equation}
     The nonzero value of the coefficient $B_2$ is due to the
contribution of four right-angled corners at the points 1 and 2.
It has been noted at first time in reference\ \cite{Dowker}. In
the case of the coefficient $B_{3/2}$ the contributions of the
corners from internal and external sectors are mutually canceled,
while the contributions of the curvature of the arc 1 -- 2 from
internal and external sectors are added.

Different geometrical origins of the zeta function failure to
provide a finite value of the vacuum energy in the two- and
three-dimensional versions of the boundary value problem in
question probably imply the impossibility of obtaining a finite
and unique value of this quantity by taking advantage of the
atomic structure of the boundary~\cite{Candelas} or its
quantum fluctuations \cite{FS1}. It is clear because any physical
reason of the Casimir energy divergences should be valid
simultaneously in the two- and three-dimensional versions of the
boundary configuration under consideration.

\section{Conclusion}
The basic result of the  paper  is  the calculation of the
individual contributions to the heat kernel coefficients generated
by such particularities of the boundary as the corners. In the
course of this analysis certain patterns, that are followed by
these contributions, have been revealed. As a by product, the
rules for obtaining all the heat kernel coefficients for the minus
Laplace operator defined on a polygon or in its cylindrical
generalization are formulated, these rules being valid both for
Dirichlet and Neumann boundary conditions. Implications of the
obtained results in calculations of the vacuum energy for regions
with nonsmooth boundary are discussed.
Our calculations comport the conventional point of view according
to which the heat kernel coefficients are determined, in the case
under consideration, by the local properties of the boundary.

     In any case, the heat kernel coefficients obtained in the
present paper can be used for  verification of the general methods
of calculating the contributions of boundary discontinuities to
the heat kernel coefficients which may be developed in the future.
These general methods must in particular allow one to calculate
the contribution of an arbitrary corner to the coefficients $B_n$
with $n>1$ in terms of the jumps of the derivatives of the
boundary curve (or its geometrical invariants) at this corner,
i.e., the formulas analogous to equation\ (\ref{eq4-5}) should be
found for $B_n$ with $n>1$.

\ack We are greatly indebted to one of the referee who found out a
substantial error in our consideration of the external sector in
the first version of the paper. This study was started  during the
visit  of VVN to Nuclear Physics Institute of Academy of Sciences
of the Czech Republic in \v{R}e\v{z} near Prague.  The work was
supported by  the Votruba-Blokhintsev Program, Russian Foundation
for Basic Research (Grant No.\ 00-01-00300), INTAS (Project No.\
578), and Grant Agency of Academy of Sciences of the Czech
Republic (Grant No.\ IAA1048101). VVN is grateful to D~V~Fursaev
for elucidating discussion of the cone singularity and its
physical implication.

\newpage

\begin{figure}
\begin{center}
\epsfbox{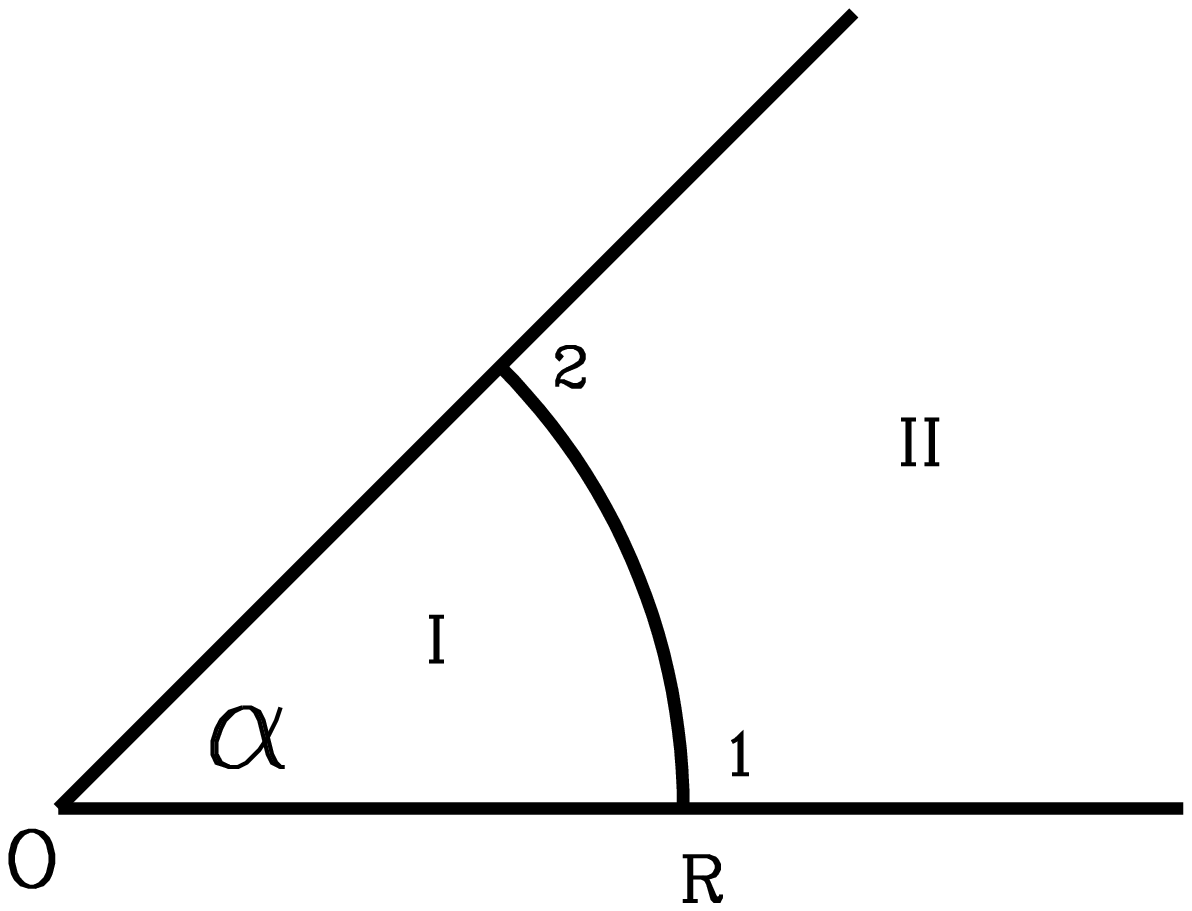}
\end{center}
\caption{\label{fig1} The cross section of a dihedral angle with
circular boundary of radius $R$ inside.}
\end{figure}

\begin{table}
\caption{\label{table1}The contributions of the functions
$Z^D_{j},\; j=-1,0, \ldots 4$ to the heat kernel coefficients
(internal circular sector I with Dirichlet boundary conditions).}
\begin{indented}
\item[]\begin{tabular}{@{}ccccccc} \br
&$Z^D_{-1}(s)$&$Z^D_0(s)$&$Z^D_1(s) $&$Z^D_2(s)$&$Z^D_3(s)
$&$Z^D_4(s)$
\\
\mr
\\
${{\displaystyle B_0}\atop (s\to 3/2)}$&$\frac{{\displaystyle 1}}
{{\displaystyle 2}}\alpha R^2$&&&&\\ \\
${{\displaystyle B_{1/2}}\atop (s\to 1)}$&$-\sqrt{\pi}\,R$&$-
\frac{{\displaystyle 1}}{{\displaystyle 2}}
\alpha R \sqrt{\pi}$&&&&\\ \\
${{\displaystyle B_1}\atop (s\to 1/2)}$&$
\frac{{\displaystyle \pi ^2}}{{\displaystyle 6 \alpha}}$&
$\frac{{\displaystyle \pi}}{{\displaystyle 2}}$&$
\frac{{\displaystyle \alpha}}{{\displaystyle 6}}$&&&\\  \\
${{\displaystyle B_{3/2}}\atop (s\to 0)}$&&&$
\frac{{\displaystyle \sqrt{\pi}}}{{\displaystyle 4R}}$&
$\frac{{\displaystyle \alpha\sqrt{\pi}}}{{\displaystyle 64R}}$&&\\      \\
${{\displaystyle B_2}\atop (s\to -1/2)}$&&&&$
\frac{{\displaystyle \pi}}{{\displaystyle 8R^2}}$&
$\frac{{\displaystyle  4}}{{\displaystyle 315}}
\frac{{\displaystyle \alpha}}{{\displaystyle R^2}}$&\\       \\
${{\displaystyle B_{5/2}}\atop (s\to
-1)}$&&&&&$\frac{{\displaystyle 25}}{{\displaystyle 96}}
\frac{{\displaystyle \sqrt{\pi}}}{{\displaystyle R^3}}$&
$\frac{{\displaystyle 37\,\sqrt \pi}}{{\displaystyle
8192\,R^3}}\,\alpha$ \\
\br
\end{tabular}
\end{indented}
\end{table}

\vspace{0.5cm}
\begin{table}
\caption{\label{table2}The contributions of the functions
$\tilde{Z}^N$ and $Z^N_{j},\; j=-1,0 \ldots 4,$ to the heat kernel
coefficients (internal circular sector I with Neumann boundary
conditions).}
\begin{indented}
\item[]\begin{tabular}{@{}cccccccc} \br & $\tilde{Z}^N(s)$
&$Z^N_{-1}(s)$&$Z^N_0(s)$&$Z^N_1(s) $&$Z^N_2(s)$&$Z^N_3(s)
$&$Z^N_4(s)$
\\
\mr
\\
${{\displaystyle B_0}\atop (s\to 3/2)}$&&$\frac{{\displaystyle 1}}
{{\displaystyle 2}}\alpha R^2$&&&&\\ \\
${{\displaystyle B_{1/2}}\atop (s\to
1)}$&$2\sqrt{\pi}\,R$&$-\sqrt{\pi}\,R$&$
\displaystyle{\frac{\sqrt{\pi}}{2}\,
 \alpha R }$&&&&\\ \\
${{\displaystyle B_1}\atop (s\to 1/2)}$&$\displaystyle{\pi}$&$
\displaystyle{\frac{\pi ^2}{6 \alpha}}$&
$\displaystyle{-\frac{\pi}{ 2}}$&$\displaystyle{
\frac{\alpha}{6}}$&&&\\  \\
${{\displaystyle B_{3/2}}\atop (s\to 0)}$&$ {
\displaystyle\frac{3\sqrt{\pi}}{2\,R}}$&&&$ \displaystyle{
-\frac{3\sqrt{\pi}}{4R}}$&
${\displaystyle \frac{5\,\alpha\sqrt{\pi}}{64R}}$&&\\   \\
${{\displaystyle B_2}\atop (s\to -1/2)}$&${\displaystyle
\frac{3\,\pi}{4\,R^2}}$&&&&$
\displaystyle{-\frac{3\,\pi}{8\,R^2}}$&
$\displaystyle{\frac{4}{45}
\frac{\alpha}{R^2}}$&\\   \\
${{\displaystyle B_{5/2}}\atop (s\to
-1)}$&$\displaystyle{\frac{21\sqrt{\pi}}{16\,R^3}}$&&&&&$\displaystyle{-\frac{
21}{32} \frac{\sqrt{\pi}}{ R^3}}$& $\displaystyle{\frac{
269\sqrt{\pi}}{ 8192}\,\frac{\alpha}{R^3}}$\\
\br
\end{tabular}
\end{indented}
\end{table}

\begin{table}
\caption{\label{table3}The contributions of different parts of the
boundary to heat kernel coefficients; $D$ and $N$ stand for the
Dirichlet and Neumann boundary conditions for a wedge,  $D_C$ and
$N_C$ denote these conditions for a cone; the upper (lower) sign
is referred to the internal I (external II) sector.}
\begin{indented}
\item[]\begin{tabular}{@{}ccccc} \br \multicolumn{2}{c}{} &
Curvature  & Right-angled corners
 & Corner of angle $\alpha$ \\
&&of the arc 1--2 & at the points 1 and 2& at the origin \\
\mr    \\
 $ B_1 $  & $D$  &$\displaystyle \pm \frac{\alpha}{3}$&$\displaystyle
\frac{\pi}{2}$&$\displaystyle {c(\alpha)\atop 0}$\\ \\
 & $N$&$\displaystyle \pm
\frac{\alpha}{3}$&$\displaystyle
\frac{\pi}{2}$&$\displaystyle {c(\alpha)\atop 0}$\\ \\
 & $D_C$&$\displaystyle \pm
\frac{\alpha}{3}$&  &$\displaystyle { 2\,
c(\alpha/2)\atop 0}$\\ \\
 & $N_C$& $\displaystyle
 \pm \frac{\alpha}{3}$&&$\displaystyle {2\, c(\alpha/2)\atop 0}$\\
\mr
$ B_{3/2}$ & $D$ &$\displaystyle
\frac{\sqrt{\pi}}{64}\,\frac{\alpha}{R}$&$\displaystyle
\pm\frac{\sqrt{\pi}}{4\,R}$&\\ \\
 & $N$ &$\displaystyle
\frac{5\sqrt{\pi}}{64}\,\frac{\alpha}{R}$&$\displaystyle
\pm\frac{3}{4}\frac{\sqrt{\pi}}{R}$&\\ \\
 & $D_C$&$\displaystyle
\frac{\sqrt{\pi}}{64}\,\frac{\alpha}{R}$&&\\ \\
 & $N_C$&$\displaystyle
\frac{5\sqrt{\pi}}{64}\,\frac{\alpha}{R}$&\\
\mr
 $ B_{2}$  & $D$ &$\displaystyle
\pm\frac{4}{315}\frac{\alpha}{R^2} $&
$\displaystyle \frac{1}{8}\frac{\pi}{R^2}$&\\ \\
 & $N$ &$\displaystyle
\pm\frac{4}{45}\frac{\alpha}{R^2}
$&$\displaystyle \frac{3}{8}\frac{\pi}{R^2}$&\\ \\
 & $D_C$ &
$\displaystyle \pm\frac{4}{315}\frac{\alpha}{R^2} $&&\\ \\
 & $N_C$ &
$\displaystyle \pm\frac{4}{45}\frac{\alpha}{R^2} $&&\\
\mr
 $ B_{5/2}$ & $D$ &
 $\displaystyle \frac{37 \sqrt{\pi}}{8192}\frac{\alpha}{R^3}
 $&$\displaystyle\pm \frac{25}{96}\frac{\sqrt{\pi}}{R^3}$&\\ \\
 & $N$&$\displaystyle \frac{269
\sqrt{\pi}}{8192}\frac{\alpha}{R^3}$&$\displaystyle
\pm\frac{21}{32}\frac{\sqrt{\pi}}{R^3}$&\\ \\
 & $D_C$ &
$\displaystyle \frac{37 \sqrt{\pi}}{8192}\frac{\alpha}{R^3} $&&\\
\\
 &$N_C$ &
$\displaystyle \frac{269 \sqrt{\pi}}{8192}\frac{\alpha}{R^3} $&&\\
\br
\end{tabular}
\end{indented}
\end{table}

\end{document}